\documentclass[a4paper,twocolumn,11pt]{quantumarticle}
\pdfoutput=1

\usepackage[utf8]{inputenc}
\usepackage[english]{babel}
\usepackage[T1]{fontenc}
\usepackage{amsmath}
\usepackage{hyperref}
\usepackage{comment}
\usepackage[normalem]{ulem}
\useunder{\uline}{\ul}{}
\usepackage{tikz}
\usepackage{lipsum}
\usepackage{braket}
\usepackage{nicefrac}
\usepackage{graphicx}
  \graphicspath{ {./images/} }
\usepackage[inline]{enumitem}
\usepackage{booktabs}
\usepackage{qsharp}
\usepackage{multirow}
\usepackage{xcolor}

    \definecolor{cud-bluish-green}{RGB}{0,158,115}
    \colorlet{comment}{cud-bluish-green!30!black!70}

\usepackage[numbers,sort&compress]{natbib}

\newcommand{\qsharp}{Q\#}

\newcommand{\tstate}{\ket{A}}

\newcommand{\affilMS}{\affiliation{Microsoft, Redmond, WA 98052, United States}}
\newcommand{\affilQTM}{\affiliation{Quantinuum, Broomfield, CO 80021, United States}}

\usepackage{algorithm}
\usepackage{algpseudocode}

    \newcommand{\linecomment}[1]{\State \(\triangleright\) {\footnotesize #1} \normalsize}

\usepackage{xcolor}   % for \textcolor
\usepackage{listings}
 \lstMakeShortInline{|}

\newcommand{\srcpath}{./src}
\newcommand{\figpath}{./figures}

\begin{document}

\title{Advances in compilation for quantum hardware -- A demonstration of magic state distillation and repeat-until-success protocols}

\author{Natalie C. Brown}
  \affilQTM
\author{John Peter Campora III}
  \affilQTM
\author{Cassandra Granade}
  \affilMS
\author{Bettina Heim}
  \affilMS
\author{Stefan Wernli}
  \affilMS
\author{Ciar\'{a}n Ryan-Anderson}
  \affilQTM  
\author{Dominic Lucchetti}
  \affilQTM
\author{Adam Paetznick}
\affilMS
\author{Martin Roetteler}
\affilMS
\author{Krysta Svore}
\affilMS

\author{Alex Chernoguzov}
  \affilQTM
\maketitle

\begin{abstract}
Fault-tolerant protocols enable large and precise quantum algorithms.
Many such protocols rely on a feed-forward processing of data, enabled by a hybrid of quantum and classical logic. 
Representing the control structure of such programs can be a challenge. 
Here we explore two such fault-tolerant subroutines and analyze the performance of the subroutines using Quantum Intermediate Representation (QIR) as their underlying intermediate representation. 
First, we look at QIR's ability to leverage the LLVM compiler toolchain to unroll the quantum iteration logic required to perform magic state distillation on the $[[5,1,3]]$ quantum error-correcting code as originally introduced by Bravyi and Kitaev  \cite{bk_uqc_2005}. 
This allows us to not only realize the first implementation of a real-time magic state distillation protocol on quantum hardware, but also demonstrate QIR's ability to optimize complex program structures without degrading machine performance. 
Next, we investigate a different fault-tolerant protocol that was first introduced by Paetznick and Svore  \cite{paetznick2013repeat}, that reduces the amount of non-Clifford gates needed for a particular algorithm. 
We look at four different implementations of this two-stage repeat-until-success algorithm to analyze the performance changes as the results of programming choices. 
We find the QIR offers a viable representation for a compiled high-level program that performs nearly as well as a hand-optimized version written directly in quantum assembly. 
Both of these results demonstrate QIR's ability to accurately and efficiently expand the complexity of fault-tolerant protocols that can be realized today on quantum hardware.  
\end{abstract}

%% MAIN BODY %%%%%%%%%%%%%%%%%%%%%%%%%%%%%%%%%%%%%%%%%%%%%%%%%%%%%%%%%%%%%%%%%

%!ROOT=../manuscript.tex

%%%%%%%%%%%%%%%%%%%%%%%%%%%%%%%%%%%%%%%%%%%%%%%%%%%%%%%%%%%%%%%%%%%%%%%%%%%%%%
\section{Introduction}
\label{sec:intro}
%%%%%%%%%%%%%%%%%%%%%%%%%%%%%%%%%%%%%%%%%%%%%%%%%%%%%%%%%%%%%%%%%%%%%%%%%%%%%%
Current quantum computing devices are fundamentally limited by noise; however, fault-tolerant protocols offer the promise of enabling arbitrarily large and precise quantum applications to run on hardware \cite{gottesmanIntroductionQuantumError2009}.
As a result, expanding the scope of quantum computing past the ``NISQ'' regime to reach broad and practical quantum advantage requires fault-tolerance as a necessary step.
With currently available quantum hardware, we can start making progress towards the goal of fault-tolerance by identifying necessary components required by common fault-tolerance proposals and developing solutions for those requirements.

In this paper, we explore two examples of fault-tolerant subroutines, both containing common elements found in quantum error correction (QEC) protocols. 
First, we implement \emph{magic state distillation} (MSD) \cite{bk_uqc_2005}, a common component to many fault-tolerance proposals, on real hardware and demonstrate how quantum programs that include classical feed-forward of QEC syndromes can help us realize magic state distillation in practice. 
Next, we study detailed \textit{simulations} of a \emph{repeat-until-success} (RUS) circuit \cite{paetznick2013repeat} used to implement arbitrary single-qubit rotations while reducing non-Clifford gate count. 
We express the circuit in a variety of logically equivalent ways and explore the trade-offs between circuit representations. 
In particular, these demonstrations show how the Quantum Intermediate Representation (QIR) allows us to represent the hybrid quantum--classical logic required for these subroutines in a way that is amenable both to analysis by compiler toolchains and to execution on current quantum hardware. 

%-----------------------------------------------------------------------------
\subsection{Fault-tolerance and magic state resources}
%-----------------------------------------------------------------------------
Practical quantum computations are inherently prone to error due to imperfect control and noise in gate operations in the hardware as well as due to unintended interactions with the environment. 
Quantum error-correcting (QEC) codes are a leading paradigm for eliminating such errors, paving the way for large-scale fault-tolerant quantum computing. 
However, universal fault-tolerant quantum computation requires a universal gate set, necessitating logical gates beyond just Clifford gates. 
No single QEC code based on qubits admits such a universal, fault-tolerant gate set as forbidden by the Eastin--Knill theorem \cite{eastin2009restrictions}.

This creates a unique set of challenges.
First, to complete a universal gate set, one must be able to produce logical gates that are not native to the QEC code, which usually comes at some overhead cost.
Second, because of this cost, ideally one must minimize the amount of resources it takes to produce these non-native logical gates.
In what follows we describe two fault-tolerant protocols that address these challenges and the demand implementing these protocols places on quantum software and hardware.  

%-----------------------------------------------------------------------------
\subsubsection{Magic state distillation: creating magic }
One of the leading proposals for circumventing the Eastin--Knill theorem utilizes a technique known as magic state distillation (MSD) and was first introduced by \citet{bk_uqc_2005}.
By first fault-tolerantly preparing a logical magic state via distillation, a logical non-Clifford gate can be implemented using a gate teleportation circuit consisting of only logical Clifford gates. 
In this way, a universal fault-tolerant gate set can be constructed from only fault-tolerant logical Clifford gates, which are well realized in most practical codes, and an MSD protocol. 

The general process of distillation takes in a number of noisy magic states, applies the unitary decoder of a QEC code, and then destructively measures syndromes of the code.
If the syndromes are trivial, a decoding circuit is used to produce one or more higher fidelity magic state(s), \textit{distilled} from the noisy input states. 
If the syndromes are non-trivial, the whole process begins again.
In general, the distillation step succeeds with a small probability; thus, multiple attempts are needed to ensure a large probability of successfully preparing a high-quality, distilled magic state $\tstate$.
In particular, distillation success is \emph{heralded}; therefore, we can discriminate between successful and unsuccessful distillation attempts via the measurements of ancilla qubits.
As such, the MSD protocol is repeated as necessary (i.e., until all syndromes are trivial) and can be represented as a quantum loop. 

\begin{figure*}[ht!]
\includegraphics[trim=15 425 250 0, clip,width=\textwidth]{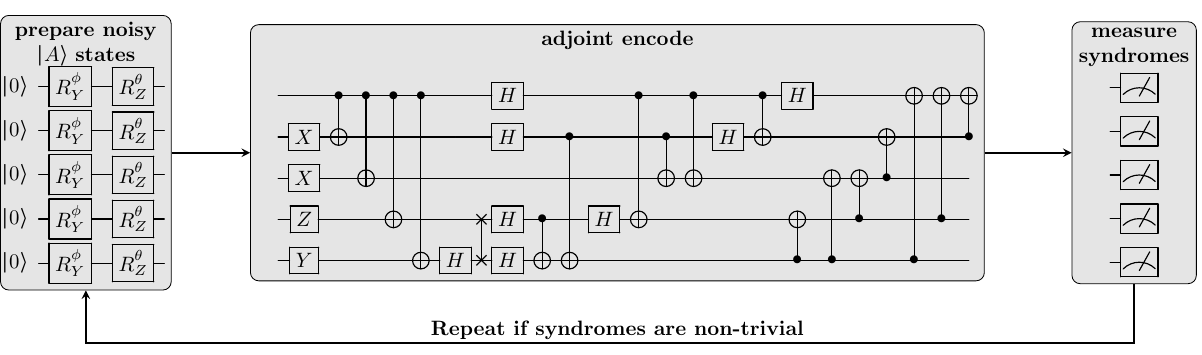}
\caption{A circuit diagram for the MSD protocol implemented. This program uses the `PrepareNoisyA` operation to prepare approximate copies of $\ket{A}$ using a decomposition into $R_Y$ and $R_Z$ operations with $\phi = \arccos(\frac{1}{\sqrt{3}})$ and $\theta = \frac{\pi}{4}$, then decodes the resulting noisy copy of
    $\ket{A} \otimes \ket{A} \otimes \ket{A} \otimes \ket{A} \otimes \ket{A}$ using the $[[5,1,3]]$ code decoder (the adjoint of the encoder). When the syndrome is observed to be 0000 (indicating no error), then this indicates that the decoded logical state is a single higher-fidelity copy of $\ket{A}$.}
\label{MSD_5_1_3}
\end{figure*}

In this work, we study the example of \citet{bk_uqc_2005} as their method is based on the $[[5,1,3]]$ quantum error-correcting code and, therefore, requires a small number of qubits to demonstrate (see Fig. ~\ref{MSD_5_1_3}). 
While in this work the $[[5,1,3]]$ MSD protocol is implemented using physical states as magic state inputs instead of logical states, it demonstrates the capability to express and carry out primitives needed to implement MSD algorithms for fault-tolerant computation.
Similar to most other MSD algorithms, the $[[5,1,3]]$ MSD method attempts to produce a high-quality, distilled magic state $\tstate$, where $\tstate\bra{A} = \frac{1}{2}(I + \frac{1}{\sqrt{3}}(X+Y+Z))$, given lower-fidelity copies. Note, $\tstate$ corresponds to the magic state Ref.~\cite{bk_uqc_2005} utilizes, which the authors of that work refer to as $\ket{T}$; however, we choose to represent the same state as $\tstate$ to avoid confusion with $T \ket{+}\bra{+}T^\dagger = \frac{1}{2}(I + \frac{1}{\sqrt{2}}(X+Y))$, where $ T = \text{diag}(1, \text{exp}(-i \pi/4))$.
Five copies of $\tstate$ are used for each distillation attempt, and distillation succeeds with probability $\nicefrac16$ such that the \citet{bk_uqc_2005} method is a 30:1 distillation protocol on average.
Realizing that 30:1 rate in practice, however, requires us to stop distillation when a successful distilled state has been obtained, such that the distillation loop cannot be unfolded into a static list of quantum instructions, but must in general include classical feed-forward, as illustrated in \autoref{alg:bk-msd}.
Moreover, since magic state distillation has to succeed on timescales much shorter than the coherence time in order to be able to synthesize $T$ operation calls in fault-tolerant systems, classical feed-forward needs to run at latency scales much smaller than the coherence time of the system.
Those requirements together imply that we must also have a way of representing the whole magic state distillation algorithm, including both its quantum and classical components. 

\begin{widetext}
  \begin{minipage}{\linewidth}
  \begin{algorithm}[H]
      \caption{Five-qubit magic state distillation with the \citet{bk_uqc_2005} protocol. }
      \label{alg:bk-msd}
      \begin{algorithmic}[1]
      \Require \text{five-qubit register \texttt{qs}}
      \Require \text{max number of attempts $N$}
      \Procedure {Distill}{}
          \Repeat
              \State \text{noisily prepare \texttt{qs} in the $\ket{A}^{\otimes 5}$ state}
              \State \text{apply the 5-qubit perfect code decoder to \texttt{qs}}
              \State \texttt{result} $\gets$ \text{measure each syndrome qubit in the $Z$-basis}
              \linecomment{\texttt{result} should be the all-zero syndrome with probability $\nicefrac16$.}
          \Until \text{\texttt{result} is the all-zero syndrome or max number of attempts exceeded}
      \EndProcedure
     \end{algorithmic}
  \end{algorithm}
  \end{minipage}
  \end{widetext}
  
%-----------------------------------------------------------------------------
\subsubsection{Repeat-until-success circuits: minimizing magic}
\label{intro:RUS}
%-----------------------------------------------------------------------------
\begin{figure*}[ht!]
\includegraphics[trim=0 450 250 0, clip,width=\textwidth]{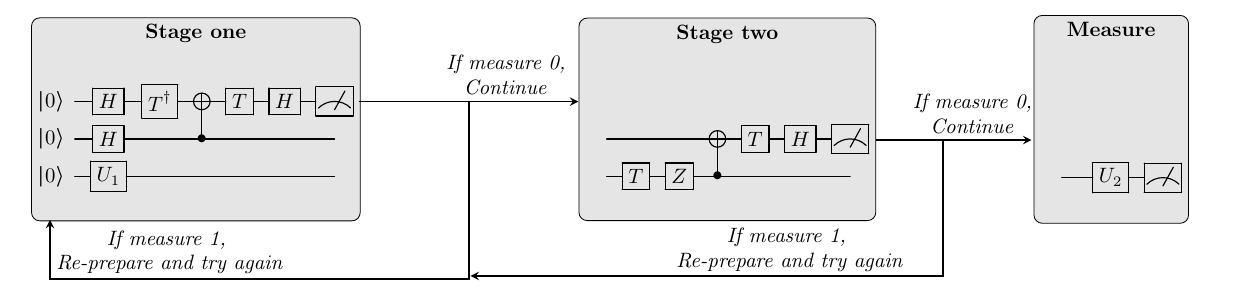}
\caption{A circuit diagram for the repeat-until-success circuit implemented. This particular repeat-until-success circuit has two criteria for success. We prepare are target qubit in some Pauli basis state by appling $U_1$ (i.e. for the $X$ basis $U_1 = H$), and the first half of the circuit executes (stage one). If the ancilla qubit is measured to be zero, the circuit continues. If not, the circuit resets and starts again. In principle, only the first ancilla need to be reset, but one could also do a hard reset where all qubits are reinitialized and reprepared. If the circuit continues, the second part of the circuit executes (stage two). Upon the second ancilla measuring zero, the success criteria is satisfied, and the quantum loop exits. If the circuit is successful, then the $V_3$ state was performed perfectly on the target qubit. We perform an inverse of unitary $U_2 = U_1^\dag V_3^\dag$ on the target qubit before measuring, as to determine the survival probability of the initial input state.}
\label{fig:RUS_cir}
\end{figure*}

Logical gate synthesis algorithms map a sequence of logical quantum operations to an equivalent fault-tolerant circuit representation by decomposing the constituent parts into a chosen native gate set (often Clifford plus $T$)~\cite{solovaykitaev, QCQI, klichnikov2013}. 
These decomposition techniques aim to approximate arbitrary unitaries within a certain error regime, where a greater number of resource states are generally needed to achieve better precision. 
The overhead associated with producing high quality magic states necessitates the need to minimize such resource states in general algorithms. 

In \cite{paetznick2013repeat}, Paetznick and Svore introduced a decomposition algorithm that approximates an arbitrary single-qubit unitary by utilizing ancilla qubits and measurements to condition subsequent gates and allows for a high level of precision with a minimal amount of non-Clifford resources.
They called these ancilla-based non-deterministic circuits "repeat-until-success" (RUS) circuits and further characterized what single-qubit unitaries could be represented exactly.
By executing certain gates conditionally on particular measurement outcomes of the ancilla, the circuit either successfully realizes the desired unitary or results in a unitary transformation that can be reversed, thus allowing the procedure to be repeated if desired. 

In this work, we study one of their RUS circuits (see \autoref{fig:RUS_cir}), which utilizes two ancilla qubits to implement the unitary $V_3 = (I + 2iZ) /\sqrt{5}$.
This circuit is a two stage RUS circuit: the first part of the circuit executes, then based on the measurement outcome, the circuit either continues or restarts. 
If the circuit continues to the second stage, then there is again a conditional measurement.
The result of this measurement determines whether or not the unitary was implemented correctly or whether the circuit must re-start again. 
Once the correct unitary has been executed, the RUS circuit is finished.
Similar to the MSD protocol, these RUS circuits can be represented as a quantum loop.       

\begin{widetext}
	\begin{minipage}{\linewidth}
		\begin{algorithm}[H]
			\caption{Repeat-until success with the \cite{paetznick2013repeat} protocol}
			\label{alg:rus}
			\begin{algorithmic}[1]
				\Require \text{three-qubit register \texttt{qs}}
				\Require \text{max number of attempts $N$}
				\Procedure {Prepare $V_3$ state on target qubit via RUS}{}
				\Repeat
					\State \text{let one of each qubit in \texttt{qs} be referred to as a auxiliary, resource, and target qubit}
					\State \text{prepare the auxiliary, resource, and target qubits in the zero state}
					\State \text{apply the appropriate two-qubit entangling unitary between the auxilary and resource qubits}
					\State \texttt{result1} $\gets$ \text{measure the auxiliary qubit in the $X$-basis}
				\If {\texttt{result1} = 0}
					\State \text{apply the appropriate two-qubit entangling unitary between resource and target qubits}
					\State \texttt{result2} $\gets$ \text{measure the resource qubit in the $X$-basis}
				\EndIf
				\Until \text{\texttt{result1} and \texttt{result2} are both zeros or max number of attempts exceeded}
				\EndProcedure
			\end{algorithmic}
		\end{algorithm}
	\end{minipage}
\end{widetext}

%-----------------------------------------------------------------------------
\subsection{Experimental realization of MSD and RUS protocols on Quantinuum devices}
%-----------------------------------------------------------------------------

To be able to execute potentially indeterminate processes such as MSD or the RUS circuits on a quantum computer introduces rather advanced requirements for the system's software and hardware. 
The $[[5,1,3]]$ MSD protocol has been demonstrated before using an NMR device \cite{souza2011experimental} but relied on post-selection as means for determining distillation success. For performance reasons, such an approach is impractical to achieve fault-tolerance for larger systems. 
A suitable choice of program representation can significantly facilitate realizing that support in a more scalable manner.
The Quantum Intermediate Representation (QIR) admits a structured representation of a program's control flow that allows for an ergonomic translation for many hardware control systems.

In this work, we demonstrate an improved experimental realization of an MSD protocol by leveraging QIR and the mid-circuit measurement and reset (MCMR) support provided by the H1 devices.
We discuss the program representation in \autoref{sec:representation}, its compilation to Quantinuum's trapped ion quantum charge-coupled device (QCCD) in \autoref{sec:compilation}, and results of our experiments for both MSD and RUS protocols in \autoref{sec:experiment}.

%!ROOT=../manuscript.tex

%%%%%%%%%%%%%%%%%%%%%%%%%%%%%%%%%%%%%%%%%%%%%%%%%%%%%%%%%%%%%%%%%%%%%%%%%%%%%%
\section{Program Representation}
\label{sec:representation}
%%%%%%%%%%%%%%%%%%%%%%%%%%%%%%%%%%%%%%%%%%%%%%%%%%%%%%%%%%%%%%%%%%%%%%%%%%%%%%

The algorithm presented in \autoref{alg:bk-msd} can be expressed in a quantum programming language like \qsharp~with a recursive function or loop (see \autoref{lst:qsharp-msd}).
While we choose to primarily work with \qsharp~in this paper based on its well developed compiler for hybrid computations, it should be mentioned that the work and discussions equally hold for 
other programming languages that are expressive enough and can compile to QIR. This should be true for languages like OpenQASM~3~\cite{crossopenqasmthree}, given enough compiler development.

The use of a high-level language or a suitable quantum programming framework abstracts backend-specific concerns from the actual application logic. This allows us to intuitively construct, compose, and test an application or algorithm in a natural manner and, thus, significantly reduces the development effort to explore novel application patterns. In the case of magic state distillation, it is, for example, desirable to use a variable to control the distillation limit rather than hand-crafting a dedicated implementation for each limit. At the same time, it is necessary to translate such an abstraction into a representation that allows hardware backend compilers to fully schedule all necessary QPU instructions. Thus, the \qsharp~program presented in \autoref{lst:qsharp-msd} needs to be \emph{compiled} into a suitable \emph{intermediate representation} before it can be translated into a native format to execute on a given QPU.  

%=============================================================================
\subsection{Intermediate representation for hybrid applications}
\label{sec:qir-motivation}
%=============================================================================

The Quantum Intermediate Representation (QIR)~\cite{QIRSpec2021} is an open specification of such a
compiler intermediate representation developed with the goal to accelerate progress, growth, and interoperability within the quantum ecosystem.
It defines how to represent quantum and classical logic using a subset of the LLVM IR~\cite{LlvmIR}.
The LLVM IR serves as a common backend-agnostic representation that a wide range of programming languages compile to~\cite{LLVMwiki}. Given a target-specification, this intermediate representation is then further optimized and translated into machine code by a code generation backend. 

Rather than being intended for humans to write or to readily read, the purpose of QIR is to be easy for compilation tools to process and optimize.
One of the advantages of using a subset of the LLVM IR to represent QPU kernels is that existing compilation techniques and tools can readily be applied to significantly simplify and optimize the program logic in hybrid quantum applications. The LLVM toolchain is used in most state-of-the art classical compilers to target a variety of processors, including CPUs, GPUs, and FPGAs~\cite{NVVMdocs, OpenMPtoFPGA}. 
Thus, QIR follows the same approach pursued to integrate GPU accelerators into larger heterogenous multi-processor systems. 

With the use of QIR we have an extensive set of mature state-of-the-art LLVM tools for program analysis and code transformation at our disposal to support the translation of a natural algorithmic representation into performant machine code. Performance optimization can come from reductions applied directly by the source
language compiler, a transformation on the intermediate representation, or architecture-specific instruction scheduling.
We omit a detailed discussion of language-specific optimizations since they are mostly irrelevant for the experiments discussed in this paper. For the remainder of this section, we briefly outline some considerations for transformations at the IR stage, and \autoref{sec:compilation} discusses some of the backend-specific compilation in more detail.

%=============================================================================
\subsection{Reduction to real-time computations}
\label{sec:profile-reduction}
%=============================================================================

\begin{figure}[]
	\begin{center}
		\includegraphics[scale=0.3]{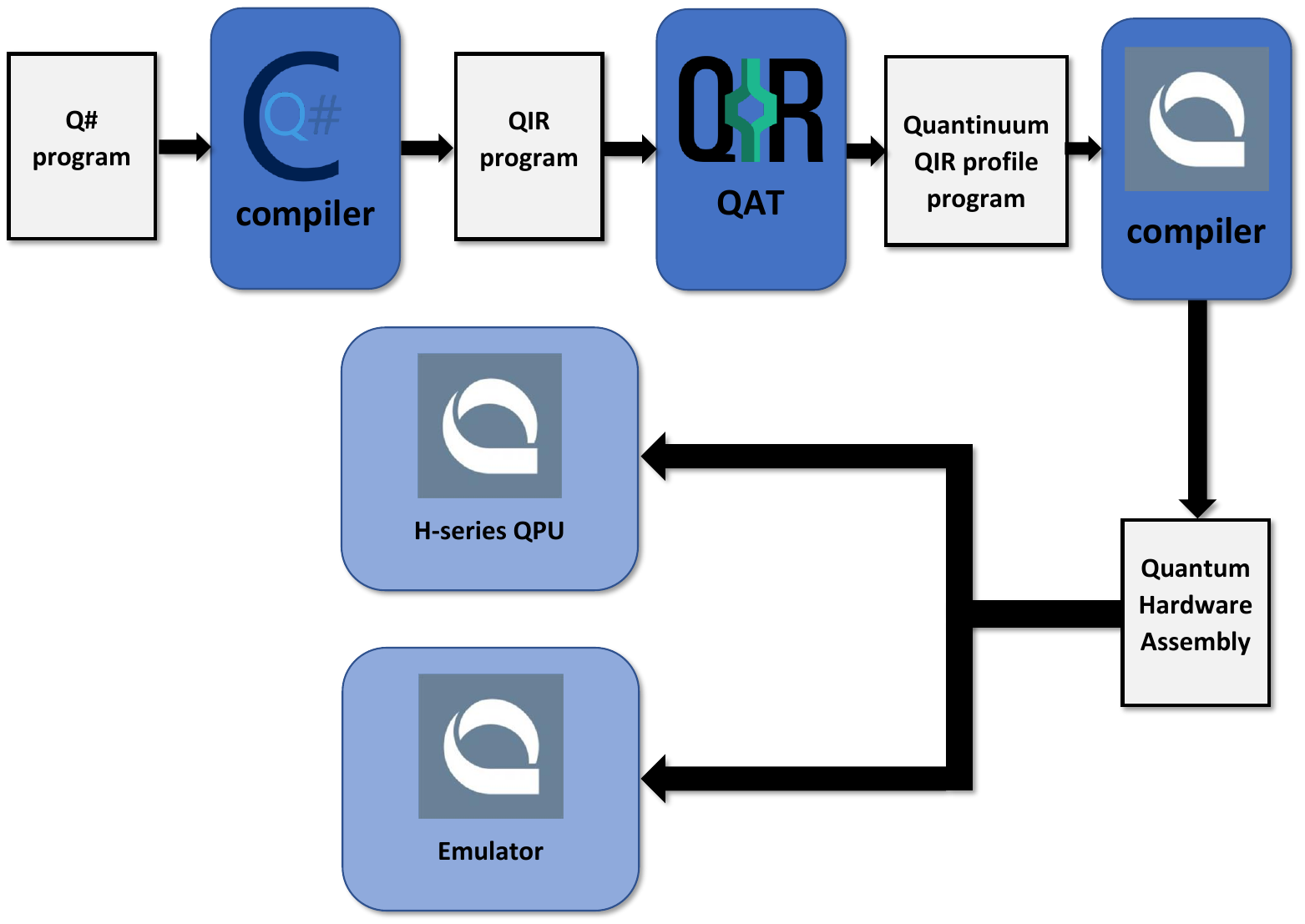}
		\caption{
			\label{fig:Compilation_Flow}
			The compilation flow of a \qsharp program. The \qsharp compiler generates a QIR program. This QIR is sent through QAT which transforms this general-purpose QIR into a target-specific profile, in this case Quantinuum's H1 series profile. This tailored QIR is then sent through the Quantinuum compiler which generates the hardware assembly and then sent to the target device. 
		}
	\end{center}
\end{figure}

As powerful as the LLVM toolset is, there are some caveats worth keeping in mind when working with quantum accelerators.
One is that quantum hardware backends tend to vary more in their architectures than CPU and GPU architectures used in personal computers, mobile devices, etc. Backend-specific compilation stages are highly customized and may not rely on existing tools for machine code generation at this time. Correspondingly, reducing complexity in the IR as much as possible prior to any backend-specific processing is a priority. For example, cycles in the call graph of a program may be challenging to process since all quantum instructions need to be statically scheduled on most currently available hardware backends. It may hence be necessary to flatten otherwise recursive structures during compilation, even if this significantly increases code size. This flattening is, for example, a necessary
transformation for running the RUS circuit on Quantinuum's QPU when evaluating a representation of the circuit in \qsharp\ using recursive functions.

Furthermore, the metrics and requirements for program optimization in the quantum case differ from those for classical targets. For example, the amount of classical computation and data exchange performed in real time during quantum execution rather than in pre- or postprocessing is limited by the coherence time of the quantum state. It is hence generally desirable to evaluate classical computations and perform constant folding during compilation when possible, even if this increases compilation time.

QIR, in its full generality, is designed to represent arbitrarily complex scenarios and permit to delegate more responsibility to the runtime in the future. As it stands today, resource and hardware constraints in practice limit how sophisticated classical computations performed in real time can be. 
To facilitate target-specific compilation and optimization for this work, we reduce the QIR representation to contain only the logic that necessarily needs to be performed in real time, as discussed in more detail in \autoref{sec:experiment}. 
The QIR specification defines a set of \emph{profiles} to formalize the expected output after this reduction (see Fig. ~\ref{fig:Compilation_Flow}). Profiles impose additional requirements on the compiled IR depending on what the targeted execution environment supports.

Rather than quantum programming frameworks emitting profile-compliant IR, the QIR toolchain intends for frontends to emit general QIR that is then targeted to a specific profile as part of a QIR compilation stage.
An open-source tool known as the QIR Adapter Tool (QAT) \cite{qat} serves that purpose and is used to transform general-purpose QIR to a subset of QIR tailored to quantum hardware such as Quantinuum's quantum computers. This transformation may fail if the general representation cannot reasonably be simplified to match the targeted profile. A simplification may be unreasonable to perform even though it is possible in principle if it introduces too much overhead to execute in practice. An example for such a case may be to use the principle of deferred measurement to eliminate the use of mid-circuit measurements for backends that do not support them.
An outline of the full compilation flow can be seen in Fig. ~\ref{fig:Compilation_Flow}. 

A more applicable example in our case for a transformation that can be done automatically but is not currently available in QAT is the translation of recursions into loops and vice versa. In our RUS experiments, we hence evaluate performance for two different versions of the source code, one using a for-loop with a conditional block in the loop body (see \autoref{lst:qsharp-rus-loop}), and the other a recursive formulation of the same logic (see \autoref{lst:qsharp-rus-recursion}). We expect that in the future, suitable transformation passes can be added to QAT to automatically compile a less performant representation in the source code into the intermediate representation that is most suited for the targeted backend.

%=============================================================================
\subsection{Adaptive Profile used for MSD and RUS experiments}
\label{sec:adaptive-profile}
%=============================================================================

While implementation efforts for a couple of profiles are in early access stages, the official specification at the time of writing is in the process of being formally defined based on our experience and not yet available in full. A formal specification for the profile that restricts the IR to require minimal support from a quantum backend can be found in Ref.~\cite{BaseProfileSpec}. We briefly summarize additional features beyond that used in this work, which are supported as part of a more advanced profile that is under development~\cite{AdaptiveProfileWorkstream} for backends that support a certain amount of real-time classical computations and control flow.

At a very basic level, qubits and measurement results are represented as opaque structs and QPU instructions are exposed as function declarations. This grants the executing backend autonomy over how to best implement quantum-specific data types and instructions for execution. 
As far as classical data types are concerned, as long as the source language compiler relies only on built-in LLVM data types and function local memory, it is possible to ensure that any use of such values will be replaced by the appropriate constants as part of compilation -- e.g., as done by QAT -- if their exact value is known at compile time, leaving only instructions for real-time computations in the code. 
The \qsharp~source code in \autoref{lst:qsharp-msd}, to give a simple illustration, contains a multiplication of floating point values in $PrepareNoisyA$ that will be evaluated and replaced by a constant rotation angle as part of the compilation to the adaptive profile we used. 
A more detailed discussion on the use of LLVM data types and compilation passes to represent and simplify more complex constructs, such as array manipulations in the source code for example, is out of scope for this paper. We instead focus our discussion on the profile representation of the real-time computations and the control flow structure in particular. The latter reflects a core requirement for the MSD and RUS subroutines we investigate and their execution performance.

The Quantinuum backend offers the following support for real-time execution:
\begin{enumerate}
  \item An arbitrary amount of virtual registers for classical integer or boolean data.
  \item Arithmetic and comparison operations on these registers performed in real time.
  \item Arbitrary 'forward' branching with blocks of interleaved quantum and classical operations, including potentially nested control flow.
  \item The ability to perform mid-circuit measurements and use the measured value in subsequent computations, including control flow statements.
  \item The ability to use qubits after measurements and reset their state if needed.
  \item Recording and returning program output including any of the classical data computed during execution.
\end{enumerate}

The MSD and RUS programs primarily take advantage of QIR's ability to express nested control flow, express jumps to later blocks in the program (skipping other blocks of quantum operations),
and to have loops unrolled and expressed as repeated series of blocks. The control flow graph for the RUS program shown in \autoref{apx:RUS_CFG} takes advantage of all of these features.
Of particular interest is the option to easily recognize potential early exits; this allows the program execution to not incur idle time or suffer from additional noise if the protocol succeeds early
in execution of the program--while still being able to execute repeated retries of operations when a protocol does not succeed.

%!ROOT=../manuscript.tex

%%%%%%%%%%%%%%%%%%%%%%%%%%%%%%%%%%%%%%%%%%%%%%%%%%%%%%%%%%%%%%%%%%%%%%%%%%%%%%
\section{QCCD compilation}
\label{sec:compilation}
%%%%%%%%%%%%%%%%%%%%%%%%%%%%%%%%%%%%%%%%%%%%%%%%%%%%%%%%%%%%%%%%%%%%%%%%%%%%%%
\newcommand{\EDSL}{EDSL}

Because QIR gives a structured representation of nested control flow that cannot be expressed in this form in more restrictive program representations,
it is important for Quantinuum's QPU to take advantage of the richer information conveyed by the representation. Moreover, in order for the MSD and RUS
circuits to take full advantage of QIR as a representation and compilation target for high-level quantum applications,
the QPU needs some way of optimizing performance based on the representation of these programs. To that end, we briefly discuss some aspects of the hardware
and control system.

\subsection{Experimental realization}

\subsubsection{Control system}
At a high level, Quantinuum's compilation infrastructure works by compiling input programs in representations like QIR to into a series of real-time
executable instructions for controlling the QPU. The compilation pipeline works by first compiling a user program (a Quantinuum profile
QIR program for example) into a program in an embedded domain-specific language (an \EDSL\ for short) that is used by experimental physicists to program hardware subroutines
on the QPU. The compilation pipeline then works by compiling this \EDSL\ program into a byte-code program that interacts with the real-time control system software stack.
In this section, we refer to the \emph{Quantinuum compiler} as the application that takes user programs in as input and produces an \EDSL\ program as output.

This \EDSL\  serves two purposes: (1) to work as an intermediary
between the Quantinuum compiler and a real-time bytecode interpreter and (2) to allow experimental physicists that develop the hardware to define the operations
emitted by the compiler (for example provide a definition for how to swap ions in some location on the trap) in a language that still has some high-level abstractions.
Quantinuum's control sytem software is controlled by a real-time bytecode interpreter that executes bytecode programs representing a shot of a quantum program.
This bytecode interpreter has operations for interacting with the lab hardware, an FPGA, etc.
Sequences of hardware interactions allow operations to be performed that cause ions to be transported around the trap into desired locations
and gates to be performed when
ions are in the correct positions in the trap.

One important aspect of the control system software and the QCCD architecture is that transport operations can be applied conditionally, along with gates.
Normally, the real-time engine performs all transport operations in the program regardless of control-flow
but only conditionally applies gates. However, for highly conditional programs like the MSD and RUS applications
explored in this paper, performing all of this extraneous transport decreases the accuracy of the results, and increases the overall time taken to execute the application.
Consequently,
it is best to have large conditional blocks containing many gates in an intermediate representation so that all of the gates and transport for the gates can be conditionally
skipped over. Skipping over both gates and transport leads to performance gains both in terms of reducing execution time and also preventing possible errors.
We refer to the mode of execution where gates and transport are both skipped over as \emph{conditional transport}. Conditional transport is one of the key reasons why the
structured control flow in QIR can allow for greater performance gains on Quantinuum's hardware.

We now briefly discuss the compilation targets at Quantinuum before moving to description of the overall compilation goals and structure with respect to these targets.

\subsubsection{QCCD hardware}

Experimental realizations for the MSD protocol are ran on Quantinuum's H1-1 system, whereas both the MSD and the RUS experiments are ran on Quantinuum's emulator.
The H1-1 system uses a surface electrode trap to control 20 $^{171}$Yb$^+$ ions, which serve as the qubit ions, and 20 $^{138}$Ba$^+$ ions, which function as sympathic cooling ions \cite{Barrett03}, in a QCCD quantum architecture ~\cite{Pino2020, Wineland98}. 
The QCCD architecture enables several crucial features for this work. 
Ion transport operations allows for a re-configurable qubit register, enabling effective all-to-all connectivity of the qubits \cite{Pino2020,Kaushal20}. 
Isolated gate zones allow for mid-circuit measurement and reset operations with low crosstalk error, essential for determining success and exit criteria for both the  RUS and MSD experiments.  
For more details on the QCCD architecture see ~\cite{Pino2020, ryan2021realization, ryan2022implementing} and for most recent measured error parameters see ~\cite{qtmspec}.

\subsubsection{Emulator}

Emulation of the device entails state-vector simulations and the application of a realistic error model based on details of ion trap physics ~\cite{pecos, crathesis, ryan2021realization, ryan2022implementing} and experimentally measured error parameters~\cite{qtmspec}. 
The detailed error model includes depolarizing gate noise, leakage errors, measurement and reset crosstalk errors, and coherent dephasing during transport and qubit idling due to magnetic field fluctuations \cite{ryan2021realization}. 
The emulation is run full-stack, meaning the same EDSL instructions that are generated by the compiler and sent to the actual ion-trap quantum computer are also sent to and interpreted by the emulator. 
This includes native gate decompositions, durations, transport operations, qubit position, ideal quantum operations, classical operations, and quantum/classical conditionals. All of this information is used by the emulator to update the classical and quantum state including feeding information to and appropriately applying the detailed error model.

\subsection{Primary Responsibilities of Compilation}
As mentioned, the primary responsibility of the Quantinuum compiler is to take input quantum program representations (currently OpenQASM 2.0++\footnote{We write OpenQASM 2.0++ because Quantinuum can take in an extended form of Open QASM 2.0. This extended form contains real-time support for classical operations in the middle of the circuit, along with conditional expressions that rely on these classical calculations that are performed in real time. These extensions and QIR classical instructions compile to the same hardware instructions discussed in \autoref{sec:classical}.} ~\cite{crossopenqasmtwo} and QIR ~\cite{QIRSpec2021})
and transform them into a program within the \EDSL. One important aspect of this \EDSL\ program is that it contains a fully resolved ion placement for a given ion-trap, 
an entirely planned out sequence of ion transportation across all control flow paths, and a resolved sequence of gates to perform.
The Quantinuum compiler also is responsible for trying to produce a program that maximizes performance. The main performance considerations are:
\begin{enumerate}
\item Optimizing qubit placement with respect to the input program by limiting the number of ion swap transportation steps.
\item Minimizing transport steps after qubit placement is optimized.
\item Maximizing the number of two-qubit gates that can be performed in parallel.
\item Reducing circuits to an equivalent circuit with better performance.
\end{enumerate}
It accomplishes all of these with a combination of graph/optimization algorithms that work either globally or over local subcircuits. Optimizing qubit placement is done by using the Sugiyama
algorithm \citep{Sugiyama:Algorithm:1981} with a heuristic for a given trap. Minimizing transport
and maximizing the number of parallel two-qubit gates for a block of quantum
operations is performed by an algorithm that searches the statespace of possible qubit permutations
and the number of transport steps to get to a permutation. Circuit reduction is implemented
via repeated traversals over a model of the circuit pattern, matching against specific gate
sequences and rewriting them.

\subsection{Supporting Classical Computation}
\label{sec:classical}
To support arithmetic and comparison operations in QIR, the Quantinuum compiler simply uses binary operators that exist in the underlying DSL. This gets compiled to an appropriate series
of ADD, SUB, MUL, etc. instructions  using common techniques for compiling arithmetic expression trees. 

Supporting
the large amount of virtual registers that can appear requires more care, as the real-time
environment only has a finite amount of classical memory locations that it can support.
The \EDSL\  has notions of global registers
that live throughout the quantum programs. Each of these global registers directly map to one of the classical memory
locations in the real-time environment.
Mapping a virtual register to a unique
register in the \EDSL\  can quickly exhaust all available registers since virtual registers cannot
be reused in a QIR program as follows from LLVM's Static Single Assignment (SSA) form.

Consequently,
a Chaitin-style register allocation algorithm that uses graph coloring \citep{registercolor} is implemented in the Quantinuum compiler so that many virtual
registers in the QIR program map to the same global register in the compiled \EDSL\  program.
In the case that register allocation cannot color the virtual registers to use an amount
supported by the real-time bytecode interpreter, then a compile-time error is raised.
Supporting user-selected results is straightforward, so we omit its discussion for brevity.

\subsection{Supporting Nested Control Flow}
Handling nested control flow also requires care. Currently, the entire compilation and execution stack was designed around
programs with only top-level conditional statements. Consequently, handling nested conditionals in QIR programs requires
flattening the control flow into an equivalent form with only top-level conditionals in the \EDSL. Thankfully, this
is possible since the \EDSL\  does have a notion of conditional \emph{blocks} and arbitrary boolean expressions.
These, respectively, map to branch instructions and comparison operations in the real-time byte-code program.
A naive flattening algorithm may try to serialize all possible control flow paths through the program and trigger only the relevant
control flow path the program takes.

This naive approach can cause an exponential blow up in program size with respect to the
number of conditional blocks. Thankfully, an existing algorithm,
if-conversion \citep{Allen:Conversion:1983}, can flatten a program in a manner that has $O(n)$ space complexity with respect to $n$ basic blocks.
There is only small overhead in the need for adding additional classical memory to track control flow and inserting boolean predicate that guard basic blocks

If-conversion works by topologically sorting the blocks of code in a control flow graph (CFG) while \emph{guarding} blocks with boolean predicates to ensure
that they respect the control flow invariants of the original CFG.
Since the QIR adhering to the Quantinuum profile has a CFG with no back edges
to form loops, the CFG is a directed acyclic graph and, thus, straightforward to topologically sort.
Figure~\ref{fig:if-conversion} gives a visualization of what if-conversion does with respect
to the control flow of a source \qsharp\ program. Notice that the nested-if in the original program
appears as fallthrough in the if-converted program but that an additional boolean expression to track
control flow is added to the condition of the if-expression.
In reality, this transformation is done
at a much later representation internal to the compiler and no transformation is done
on source programs.

\begin{figure}[t!]
\begin{lstlisting}[language=qsharp,gobble=4,basicstyle=\scriptsize\ttfamily]
    operation BeforeIfConversion()
    : Unit {
        use q = Qubit();
        h(q);
        let r = MResetX(q);
        if (r == One) {
          use q1 = Qubit();
          h(q1);
          let r1 = MResetX(q1);
          if (r1 == One) {            
            Rz(PI(), q1);
          }
        }
        else {
          Ry(PI(), q);
        }
    }
\end{lstlisting}

\begin{lstlisting}[language=qsharp,gobble=4,basicstyle=\scriptsize\ttfamily]
    operation AfterIfConversion()
    : Unit {
        use q = Qubit();
        h(q);
        let r = MResetX(q);
        mutable r1 = 0;
        mutable cond = 0;
        if (r == One) {
          set cond = 1;
          use q1 = Qubit();
          h(q1);
          set r1 = MResetX(q1);          
        }
        else {
          Ry(PI(), q);
        }
        if (cond and r1 == One) {            
            Rz(PI(), q1);
        }
    }    
\end{lstlisting}
\caption{An illustration of if-conversion using a \qsharp\ program. The top program represents the
  control flow of a trivial \qsharp\ program, while the bottom program illustrates the control flow
  of the program as if it went through if-conversion at the source level.}
\label{fig:if-conversion}
\end{figure}

After if-conversion is performed, the remaining responsibility of the compiler is to use
a specific form of conditional block in the EDSL that informs the real-time engine to use conditional transport.
For the RUS programs, after if-conversion the conditional blocks become less likely to become taken,
and, thus, the program can quickly skip over blocks until encountering the exit block by using conditional transport
as indicated by the EDSL program.

Once the compiler has performed ion transport planning, if-conversion, and register allocation,
code generation produces an EDSL program representing an executable realization of the user program in the form of
defined gate and transport operations. After \EDSL\ compilation, the final byte-code program is produced.
Next, we discuss the MSD and RUS experiments and perform some analysis in Section~\ref{sec:experiment}.

%!ROOT=../manuscript.tex

%%%%%%%%%%%%%%%%%%%%%%%%%%%%%%%%%%%%%%%%%%%%%%%%%%%%%%%%%%%%%%%%%%%%%%%%%%%%%%
\section{Experiments}
\label{sec:experiment}

%%%%%%%%%%%%%%%%%%%%%%%%%%%%%%%%%%%%%%%%%%%%%%%%%%%%%%%%%%%%%%%%%%%%%%%%%%%%%%
\subsection{Magic State Distillation}
\label{sec:msd-experiment}
\begin{figure}[]
	\begin{center}
		\includegraphics[width=\columnwidth]{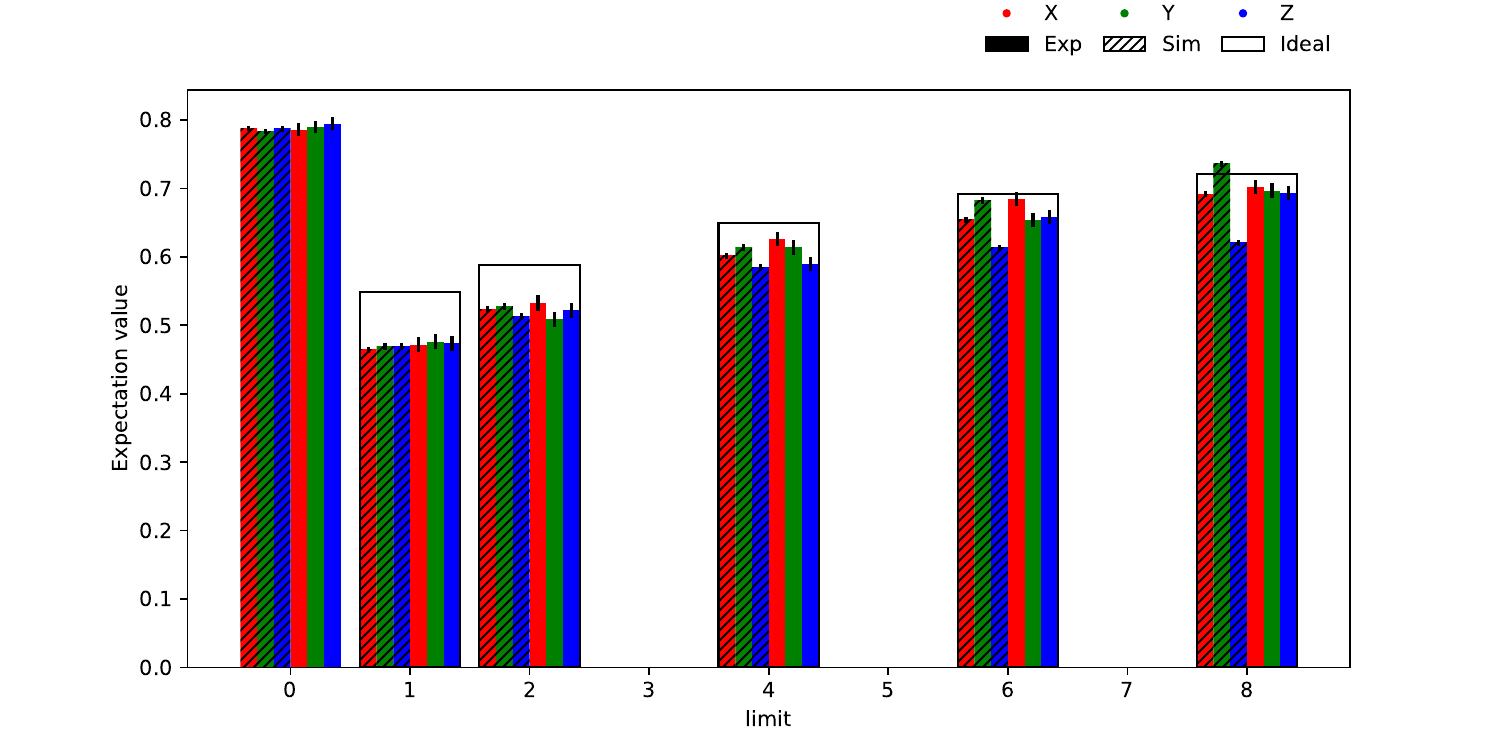}
		\caption{
			\label{fig:msd_expect_nopost}
			Expectation value along each axis of the final state prepared by magic state distillation at different limits, on the H1-1 system and the H1-1E emulator. A maximum limit of 0 indicates no distillation; that is, a raw measurement of the undistilled $\ket{A}$-state.
			Outlines in black indicate the expectation value that we would obtain from magic state distillation on a noiseless quantum device.
		}
	\end{center}
\end{figure}

\begin{figure}[]
	\begin{center}
		\includegraphics[width=\columnwidth]{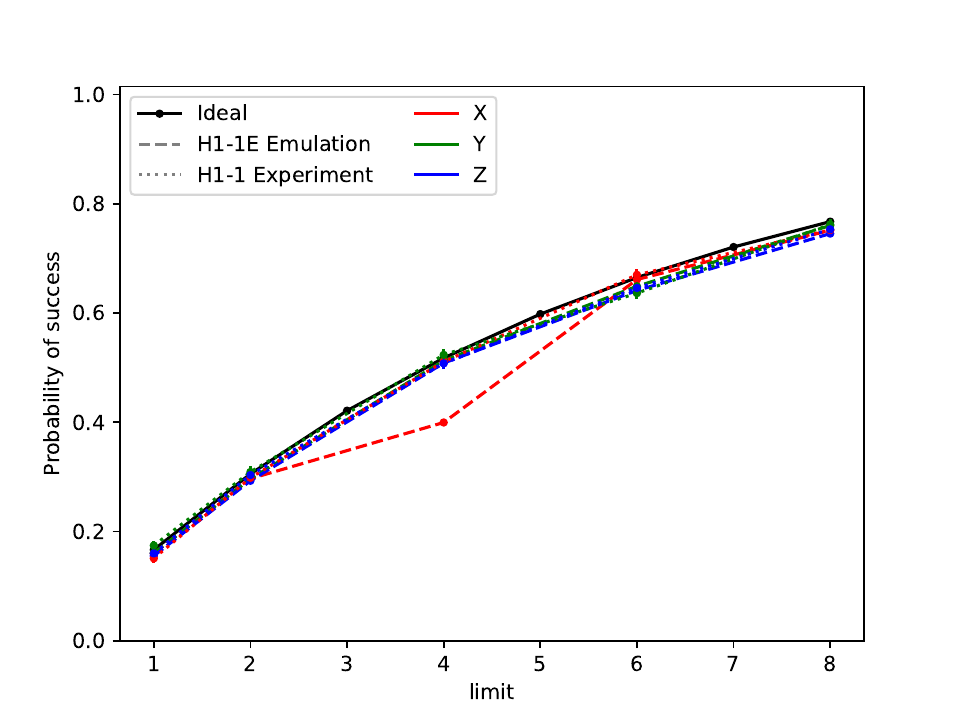}
		\caption{
			\label{fig:msd_success}
			Probability of distillation succeeding at different limits, and running both on the H1-1 system and the H1-1E emulator. The dashed black line indicates the probability of success expected for a perfect state preparation on a noiseless device; as we have intentionally chosen a state preparation procedure with systematic errors, the success probability is not expected to saturate this bound.
		}
	\end{center}
\end{figure}
To demonstrate the correctness of our representation of hybrid quantum programs, we implemented \autoref{alg:bk-msd} as a \qsharp~program (shown in \autoref{lst:qsharp-msd}) with input parameters for the final measurement basis, and for the maximum number of distillation attempts $N$, which we refer to as the limit. 

The \qsharp~program is compiled to QIR using the Quantum Development Kit\footnote{Note that this experiment used an early prerelease version of Quantum Development Kit support for compiling \qsharp~to QIR compatible with Quantinuum systems, hence the exact conventions and signatures used in our experiment may differ from those available in final versions.}.
Using the Azure CLI, this initial QIR representation is submitted to the Azure Quantum Service, reduced to an adaptive profile representation using QAT, and finally submitted to the Quantinuum backend for the target-specific compilation performed by the QCCD compiler before its execution on the the H1-2 system or its emulator.
We choose a variety of different measurement bases and limits.
For each distinct measurement basis and distillation limit, we used an entry point of the form shown as a separate Azure Quantum job:

\begin{lstlisting}[language=qsharp,gobble=4,basicstyle=\scriptsize\ttfamily]
    @EntryPoint()
    operation MeasureDistilledTAtDepth2InX()
    : Unit {
        use q = Qubit();
        PrepareDistilledT(2, q);
        let r = MResetX(q);
    }
\end{lstlisting}
For this experiment, we run 10,000 shots on the emulator and 2000 shots on hardware repectively, for limits $N$ =0, 1, 2, 4, 6 and 8. 
For each different measurement axis, we furthermore compare to a measurement of the noisy $\ket{A}$ state prepared without distillation.
A limit of $N$ = 0 indicates no distillation; that is, a raw measurement of the undistilled $\ket{A}$-state.
In order to estimate the fidelity of the distilled $\ket{A}$ state, we measure in the three respective Pauli basis $X$, $Y$, and $Z$ after a certain maximum number of attempted distillation rounds.
Given the state $\tstate\bra{A} = \frac{1}{2}(I + \frac{1}{\sqrt{3}}(X+Y+Z))$, the ideal expectation values of the different bases should all be identical.

In~\autoref{fig:msd_expect_nopost} we plot the expectation values for the actual hardware H1-1 experiment, the simulation of the H1-1 emulator which we denote as H1-1E, and the ideal results for the experiment.
In particular, we plot the expectation value regardless of whether distillation was heralded to be successful or not (where each shot has a $\nicefrac16$ probability of succeeding).
Even without post-selecting on successful distillation attempts, with increasing distillation limit the fidelity of the outcome states approach the ideal, noiseless result. 
As the limit increases, more and more shots are successful, as there are more trials for success--resulting in an overall higher average fidelity.
This relationship, between the distillation limit and probability of success can be seen in ~\autoref{fig:msd_success}.

\begin{figure}[]
	\begin{center}
		\includegraphics[width=\columnwidth]{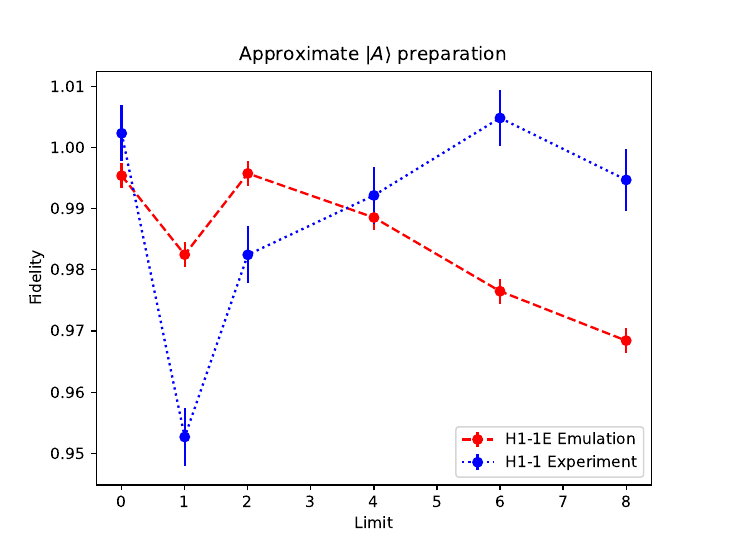}
		\caption{
			\label{fig:msd_fidelity}
			The fidelities of the final state prepared by magic state distillation at a variety of different limits, and running both on the H1-1 system and the H1-1E emulator, post-selected on successful distillation. A limit of 0 indicates no distillation; that is, a raw measurement of the undistilled $\ket{A}$-state. The figure shows the data for the approximate preparation of the $\ket{A}$ state. Since preparing this state requires only a few single qubit gates, we see the distillation procedure fail to produce a higher fidelity state. This is to be expected given the circuit depth of the procedure.}
	\end{center}
\end{figure}

In~\autoref{fig:msd_fidelity}, the state fidelity of $\ket{A}$ is plotted, post-selected on failed distillation attempts.
The number of successful shots constituting this data can been seen in ~\autoref{tab:n-msd-succ-shots}.
This shows something similar to ~\autoref{fig:msd_success}; as distillation limit grows, fewer failed shots are thrown out.
We see that the fidelity decreases due to the distillation procedure. This is expected. 
The original proposal of the $[[5,1,3]]$ MSD protocol from Bravyi and Kitaev assumed the protocol been done at the logical level (i.e., supported on a QEC code). 
Thus, the noise threshold estimates given in that paper only assumed noise on the input states, whereas the procedure is assumed to be implemented perfectly. 
In reality, running on current hardware is far from perfect. 

Furthermore, at the physical level, we would not expect a protocol consisting of many two-qubit gates like the MSD protocol to offer an advantage over a simple preparation procedure of a few single-qubit gates, as the gate error rates for two-qubit gates are typically much higher than single-qubit gates. 

We see emulated data matches relatively closely to experimental results, indicating the noise model is a reasonable representation of the noise experienced in the hardware. These results indicate that the goal of representing and executing more advanced QEC primitives utilizing QIR was achieved.

\begin{table}[]
\begin{center}
\begin{tabular}{ccc}
\hline
 \multicolumn{3}{l}{Avg. Number of Successful shots} \\ \hline
N & Sim                    & Exp.                  \\  \hline \hline
1 &1587 (16\%)              & 324 (16\%)                     \\ \hline
2 &2953 (30\%)               & 607 (30\%)              \\ \hline
4 &4737 (47\%)               & 1028 (51 \%)               \\ \hline
6 &6506 (65\%)               & 1301 (65\%)                    \\ \hline
8 &7524 (75\%)               & 1515 (75\%)             \\ \hline
\end{tabular}

\caption{
 \label{tab:n-msd-succ-shots}
The number of successful shots after post-selecting on heralded success. The total number of attempted shots for was 10000 and 2000 for the simulations and experimental runs respectively. The percentage of successful shots are in good agreement and can be seen in~\autoref{fig:msd_success}.}
\end{center}
\end{table}

%%%%%%%%%%%%%%%%%%%%%%%%%%%%%%%%%%%%%%%%%%%%%%%%%%%%%%%%%%%%%%%%%%%%%%%%%%%%%%%%%%%%%%
\subsection{Repeat-until success}
\label{sec:rus-experiment}

\begin{figure}[]
		\includegraphics[width=\columnwidth]{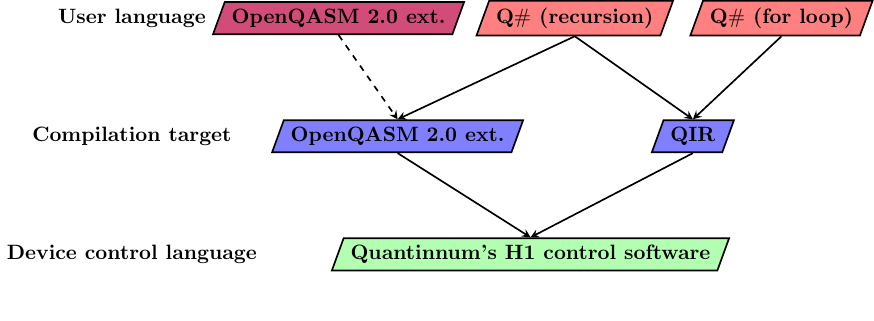}
		\caption{A schematic to show the different compilation structures of the RUS circuit implementations. The top row (red) representing what language the user wrote the circuit in. The middle row (blue) shows the compilation target. Note that for OpenQASM 2.0++ the user language and the compilation target are the same. We label this space case a different color (purple) to emphasize this. Finally, the respective IRs get compiled down according to a device profile, in this case Quantinuum H1 series.}
			\label{fig:RUS_chart}
			
\end{figure}

Next, we compare different representations of the same quantum circuit to examine performance changes due to the changes in program structure.
We simulated four different versions of the RUS circuit (\autoref{fig:RUS_cir} and \autoref{alg:rus}), each having a unique realization of the RUS program.
Each representation has a different set of language primitives that express how the algorithm should be executed. 
Changes in representation at the high-level affect the ease with which one can correctly prototype an algorithm and then refactor it to have better performance.
On the other hand, changes in representations at lower levels affect the ease and performance of compilation to control system instructions for the hardware. 
We summarize the construction of these implementations as follows: 
\begin{enumerate}
\item OpenQASM 2.0++: written in OpenQASM 2.0 and using Quantinuum's QASM extensions (\autoref{ext_qasm})
\item \qsharp~to QIR - for loop: written in \qsharp~utilizing a for loop and compiled to QIR  (\autoref{lst:qsharp-rus-loop} and \autoref{lst:rus-qir})
\item \qsharp~to QASM - recursion: written in \qsharp~utilizing recursion and automatically translated to OpenQASM 2.0++ (\autoref{lst:qsharp-rus-recursion} and \autoref{Q_qasm})
\item \qsharp~to QIR - recursion: written in \qsharp~utilizing recursion and compiled to QIR (\autoref{lst:qsharp-rus-recursion})
\end{enumerate}
Note that the last two implementations are derived from the same \qsharp~code as seen in ~\autoref{lst:qsharp-rus-recursion}. The compilation flow of these variations
are further described by \autoref{fig:RUS_chart}.

At the user level, the circuits are written in some high-level quantum language, like OpenQASM or \qsharp. 
The user then specifies what IR they wish to compile to. In the case of OpenQASM 2.0++ (or OpenQASM 2.0), this compilation target is the same as the user language. 
Finally, the user decides what devices they wish to run on; often users might tailor their code with particular devices and features in mind (i.g. mid-circuit measurement and reset). 
The culmination of these choices can affect the overall performance of the algorithm.   

The variations of the RUS circuit implementation try to capture this overall process. 
The OpenQASM 2.0++ is hand-written and optimized leveraging directly the extended QASM features available on the H1 system, and it does not rely on a high-level compiler to help with the optimization. 
It is essentially equivalent to the to a QIR program written by hand, and as such, one would expect good overall performance, at the up front cost of writing tedious code by hand. 
The \qsharp~implementations capture different coding choices within the same language and utilize a high-level compiler (see \autoref{fig:Compilation_Flow}) for optimization. 
The use of a for loop or recursion are both valid ways of representing the circuit's logic, but will get compiled to an IR in different ways. 
Finally, looking at the underlying IR and the way in which the IR gets compiled into EDSL which ultimately governs the motion of ions in the trap, will have direct implications on execution.

\begin{figure}[]
		\includegraphics[trim=15 10 20 15, clip, scale=0.75,width=\columnwidth]{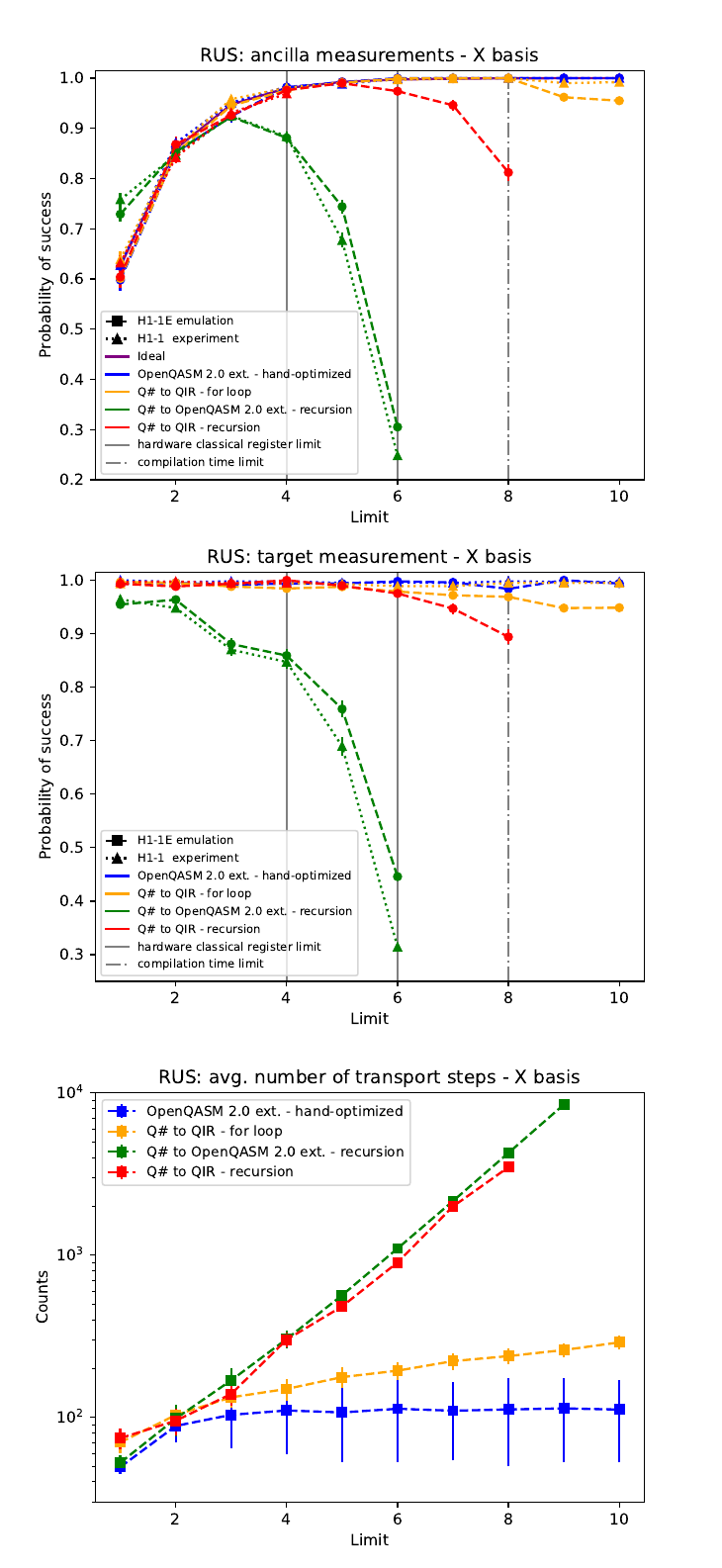}
		\caption{\textbf{a)} The performance of each program in noise-modeled emulation as retry limit is increased compared to the ideal, where success is heralded by the final measurement of the ancilla qubits used in performing the $V_3$ rotation. \textbf{b)} The probability of success as indicated by the final measurement of the target qubit, post-selected on successful ancilla qubit measurements. \textbf{c)} The average number of ion transport steps in each program. Matching graphs for Pauli-Y and Pauli-Z basis are shown in \autoref{RUS_YZ}.}
			\label{fig:RUS_X_basis}
			
\end{figure}

To analyze the performance of these representations we ran a series of simulations and experiments on both the H1-1E emulator and H1-1 quantum computer. 
For each experiment, there is a set of input parameters, the preparation and measurement basis, and the limit (i.e., the maximum number of attempts to try the RUS circuit). 
We first prepare the target qubit (i.e., the target we wish to implement the $V_3$ unitary on) in one of the three Pauli bases as designated by the input parameter.
We then perform the RUS circuit with the logic and flow structure as described in Sec. \ref{intro:RUS}.

The circuit executes according the input limit, with the circuit exiting the loop logic upon the successful measurement of the two ancilla qubits in the $\ket{0}$ state, which follows a geometric probability distribution as outlined in \cite{paetznick2013repeat}. 
With the success criteria achieved, we implement the inverse unitary $V_3^{-1}$, and measure the target qubit in the designated input basis.
In this way, we should see the same state measured out as we prepared with high fidelity.
We plot this survival probability along with the results of the simulations in \autoref{fig:RUS_X_basis} \textbf{a)}, and for the survival fidelity of the target qubit \autoref{fig:RUS_X_basis} \textbf{b)} for the $X$ basis.
Additional results for $Y$ and $Z$ bases can be seen in \autoref{RUS_YZ}. \footnote{For the higher limits of 9 and 10, the recursion implementations increased in size to the point where they were not able to be compiled for execution on the emulator. Additionally, there are limits on the number of registers a quantum program can use when executing on hardware and therefore these higher limit programs cannot execute on actual hardware, even without the compilation bottleneck.}

Initially, there is not a clear difference in performance for the program variations. 
At limits $N \leq 5$, the implementations perform roughly the same. 
However, as the limit is increased, there is a clear drop in the performance for the recursion implementations, while the OpenQASM 2.0++ and QIR - for loop variations closely track to
the expected ideal. Filtering to just the shots where the final ancilla measurements indicate success, we see the recursion implementations show lower survival fidelity of target qubit at the same 
higher limits.

To better understand this behavior, we looked at the number of transport steps each program produced.
Transport has two effects on trapped ion qubits; it adds additional memory errors and causes ion heating. 
To combat the ion heating, additional time is required to cool the qubits back to the ground state before performing gate operations.
The H1 devices are calibrated to remove extra heating accumlated during transport sequences prior to gating, but the procedure is not perfect and some residual heating may remain. 
Thus, in general, transport adds noise to the qubits, therefore more transport steps lead to degraded results.
In \autoref{fig:RUS_X_basis} \textbf{c)} we plot the relationship between the average number of transport steps and the limit. 
The recursive form has a sharply increasing average number of transport steps in both \qsharp~to QASM and to QIR, while the for loop form implemented by the \qsharp~to QIR
compilation more closely follows the trend of the hand-optimized OpenQASM 2.0++. 
The critical difference in the recursive program compilation that leads to this decrease in performance is tied to the complexity of the resulting CFG.
This is similar to how programs using loops in imperative languages often outperform an equivalent representation using recursion.

For QIR compilation, standard LLVM tools provide utilities for visualizing control flow (see \autoref{apx:RUS_CFG}).
This shows how the recursive form of the algorithm leads to higher complexity in the CFG, where size increases quadratically with chosen limit. 
Meanwhile, the for-loop form maintains a simpler CFG that only grows linearly with chosen limit. 
As described in \autoref{sec:compilation}, the QCCD compilation analyzes CFG to enable conditional transport, and the more complex CFG generated by the recursion
leads to higher transport cost during execution. 
Thus the choice of pattern at the high level implementation affects the intermediate structure in a way that allows for tailoring execution to minimize transport steps and achieve better results at higher limits.

The number of transport steps also explains the difference we see in the performance between the different bases.
We model errors due to transport as dephasing memory errors, which have a great effect on the $X$ and $Y$ basis, but only a small effect on the $Z$ basis. 
This is why we see a more dramatic drop in performance with those states when compared to the $Z$ basis.
More details of the emulator and its error model can be found in \cite{ryan2021realization, ryan2022implementing}.

These demonstrations show the importance of circuit implementation.
Writing optimized circuits by hand, tailored to a particular backend, may not be a feasible and scalable way to program a quantum computer if quantum applications develop in complexity similar to modern classical HPC or scientific computing  applications. 
Most applications in those spaces are not written in assembly languages but use compiled high-level languages that leverage programming frameworks and ecosystems that provide many abstractions over assembly languages.
It thus seems more likely that quantum application developers will eventually want to program in high-level languages with quantum abstractions, like \qsharp , and have high-level compilers and compilation techniques that optimize performance for you, analogous to classical applications.
Of course you do not want to sacrifice performance for practicality and efficiency.
By using existing compilation techniques borrowed from LLVM, QIR allows programmers to utilize abstractions that make application development easier while letting compilers generate code that can perform comparably to a hand-optimized circuit--and typically with less effort. The ability to write the RUS circuit as naturally expressed via a subroutine using a
for loop in \qsharp, and get good performance, is good support of this claim.

%!ROOT=../manuscript.tex

%%%%%%%%%%%%%%%%%%%%%%%%%%%%%%%%%%%%%%%%%%%%%%%%%%%%%%%%%%%%%%%%%%%%%%%%%%%%%%
\section{Conclusion}
\label{sec:conclusion}
%%%%%%%%%%%%%%%%%%%%%%%%%%%%%%%%%%%%%%%%%%%%%%%%%%%%%%%%%%%%%%%%%%%%%%%%%%%%%%
Quantum systems are inherently fragile.
To push past the NISQ era to a regime where quantum computers are solving meaningful problems will require at least some application of fault-tolerant protocols. 
These protocols will require real-time control of hardware, requiring quantum and classical logic.
Furthermore, large-scale algorithms will require many qubits and thus expressing algorithms in a way that is user friendly and debuggable will become increasingly more important as hardware develops.
Consequently, it seems likely that adapting approaches used to scale classical software applications will be useful in accelerating the development of such algorithms.

In this paper, we have presented an important step towards realizing fault-tolerant protocols by demonstrating that the necessary program structures for MSD and RUS type algorithms can be sufficiently analyzed and optimized by a compiler toolchain leveraging QIR to achieve the expected performance. 
In particular, we have shown that the compiler optimizations allow for the repetition of a gate sequence until the success criterion is met, without introducing additional noise due to conditionally performed gate operations that would be needed if the circuit's iterations were expressed as a static list of quantum instructions.
We also have shown that QIR provides users with a scalable and friendly solution to programming such hybrid algorithms without hindering performance. 
These demonstrations mark an important milestone in the expressivity and realization of fault-tolerant quantum circuits, notably being the first real-time demonstration of a magic state distillation protocol on real hardware. 

As quantum hardware continues to scale up and becomes increasingly more complex, the software used to represent quantum and classical logic will also need to evolve.  
QIR is a viable path forward to meet this demand, being a building block for scalable solutions. 
We expect the IR of the future to be similar to QIR, utilizing the wealth of existing classical compilation tool-chains and techniques, to enable the precise control required by fault tolerance, to harness the power of such complex quantum machines.

\begin{acknowledgements}
    This manuscript was prepared using \verb+quantumarticle+ version v\quantumarticleversion.
    This project was prepared using a reproducible workflow \cite{granade_reproducible_2017}.
    All hardware experiments were ran on the Quantinuum system model H1-2 powered by Honeywell. 
    We thank Dave Hayes and Ben Criger for useful feed-back on the manuscript and Charlie Baldwin and Karl Mayer for useful discussions.
    This work was supported by the Quantum Science Center, a National Quantum Information Science Research Center of the DOE.
\end{acknowledgements}

\bibliographystyle{plainnat}
\bibliography{bibliography}

%% APPENDIX %%%%%%%%%%%%%%%%%%%%%%%%%%%%%%%%%%%%%%%%%%%%%%%%%%%%%%%%%%%%%%%%%%

\onecolumn\newpage
\appendix
%!ROOT=../manuscript.tex

%%%%%%%%%%%%%%%%%%%%%%%%%%%%%%%%%%%%%%%%%%%%%%%%%%%%%%%%%%%%%%%%%%%%%%%%%%%%%%
\section{Source Code}
\label{apx:example}
%%%%%%%%%%%%%%%%%%%%%%%%%%%%%%%%%%%%%%%%%%%%%%%%%%%%%%%%%%%%%%%%%%%%%%%%%%%%%%
\subsection{Magic state distillation}
\begin{lstlisting}[label={lst:qsharp-msd},language=qsharp,caption={\qsharp~program implementing \autoref{alg:bk-msd}. Note that code was modified slightly for clarity from the original version. Here we named the subroutines "PrepareNoisyA", 
"DistillA", etc. as to not confuse the reader with the alternative definition for $\ket{T}$ that appears in the RUS algorithm. These changes are cosmetic and do not affect the algorithm.},escapeinside={(<}{>)},basicstyle = \ttfamily\tiny]
namespace Magic_State_Distillation {
    open Microsoft.Quantum.Intrinsic;
    open Microsoft.Quantum.Canon;
    open Microsoft.Quantum.Measurement;

    // NB: We use the definitions in (<\citet{bk_uqc_2005}>)
    // for each type of magic state.

    /// # Summary
    /// An example entry point for testing distilled A in the Pauli-X basis with
    /// maximum depth 2.
    /// Other entry points are eliminated for brevity.
    @EntryPoint()
    operation MeasureDistilledTAtDepth2InX() : Result {
        use q = Qubit();
        PrepareDistilledA(2, q);
        return MResetX(q);
    }

    operation PrepareDistilledA(maxDepth : Int, target : Qubit) : Unit {
        use aux = Qubit[4];
        Distill(maxDepth, target, aux);
        // Distilling with the (<\citet{bk_uqc_2005}>) protocol flips
        // (<$\ket{A_0}$>) to (<$\ket{A_1}$>), so we undo that here.
        H(target);
        Y(target);
    }

    /// # Summary
    /// Prepares a noisy (<$\ket{A}$>)-type magic state.
    /// This preparation is intentionally approximate
    operation PrepareNoisyA(target : Qubit) : Unit {
        Ry(2.0 * 0.4776583090622547, target); // == arccos(1/sqrt(3))
        Rz(2.0 * -2.748893571891069, target); //pi/4
    }

    operation Distill(
        maxNIterations : Int,
        target : Qubit,
        aux : Qubit[]
    ) : Unit {
        PrepareNoisyA(target);
        for q in aux {
            PrepareNoisyA(q);
        }

        Adjoint Encode(target, aux[0], aux[1], aux[2], aux[3]);
        let (meas0, meas1, meas2, meas3) = 
            (MResetZ(aux[0]), MResetZ(aux[1]),
            MResetZ(aux[2]), MResetZ(aux[3]));

        if maxNIterations > 1 {
            if AnyNonZero(meas0, meas1, meas2, meas3) {
                Reset(target);
                return Distill(maxNIterations - 1, target, aux);
            }
        }
    }

    /// # Summary
    /// Specialized test to effeciently check multiple results on hardware.
    /// Equivalent to `r0 == One || r1 == One || r2 == One || r3 == One`.
    operation AnyNonZero(
        r0 : Result,
        r1 : Result,
        r2 : Result,
        r3 : Result
    ) : Bool {
        body intrinsic;
    }

    /// # Summary
    /// The coherent encoder for the the five-qubit perfect code.
    operation Encode(
        q0 : Qubit,
        q1 : Qubit,
        q2 : Qubit,
        q3 : Qubit,
        q4 : Qubit
    ) : Unit is Adj + Ctl {
        CNOT(q1, q0);
        CNOT(q3, q0);
        CNOT(q4, q0);
        CNOT(q2, q1);
        CNOT(q3, q2);
        CNOT(q4, q2);
        CNOT(q4, q3);
        H(q0);
        CNOT(q0, q1);
        H(q1);
        CNOT(q0, q2);
        CNOT(q1, q2);
        CNOT(q0, q3);
        H(q3);
        CNOT(q1, q4);
        CNOT(q3, q4);
        H(q0);
        H(q1);
        H(q3);
        H(q4);
        SWAP(q3, q4);
        H(q4);
        CNOT(q0, q4);
        CNOT(q0, q3);
        CNOT(q0, q2);
        CNOT(q0, q1);
        X(q1);
        X(q2);
        Z(q3);
        Y(q4);
    }
}
\end{lstlisting}
\subsection{Repeat until success}
\begin{lstlisting}[label={lst:qsharp-rus-recursion},language=qsharp,caption={\qsharp~program implementing the circuit in \autoref{fig:RUS_cir} using recursion.},escapeinside={(<}{>)},basicstyle = \ttfamily\tiny]
namespace Microsoft.Quantum.Samples {
    open Microsoft.Quantum.Intrinsic;
    open Microsoft.Quantum.Canon;
    open Microsoft.Quantum.Math;
    open Microsoft.Quantum.Preparation;
    open Microsoft.Quantum.Diagnostics;

    /// # Summary
    /// Example of a Repeat-until-success algorithm implementing a circuit
    /// that achieves exp(i*ArcTan(2)*Z), also known as the "V gate"
    /// by Paetznick & Svore.
    ///
    /// # References
    /// - [ *Adam Paetznick, Krysta M. Svore*,
    ///     Quantum Information & Computation 14(15 & 16): 1277-1301 (2014)
    ///   ](https://arxiv.org/abs/1311.1074)
    /// For circuit diagram, see file RUS.png.
    ///
    /// # Input
    /// ## inputValue
    /// Boolean value for input qubit (true maps to One, false maps to Zero)
    /// ## inputBasis
    /// Pauli basis in which to prepare input qubit
    /// ## limit
    /// Integer limit to number of repeats of circuit
    ///
    /// # Remarks
    /// The program executes a circuit on a "target" qubit using an "auxiliary"
    /// and "resource" qubit. The circuit consists of two parts (red and blue
    /// in image).
    /// The goal is to measure Zero for both the auxiliary and resource qubit.
    /// If this succeeds, the program will have effectively applied an
    /// Rz(arctan(2)) gate (also known as V_3 gate) on the target qubit.
    /// If this fails, the program reruns the circuit up to <limit> times.
    @EntryPoint()
    operation AllocateQubitsAndApplyRzArcTan2 (
        inputValue : Bool,
        inputBasis : Pauli,
        measurementBasis : Pauli,
        limit : Int
    )
    : Result[] {
        
        use (auxiliary, resource, target) = (Qubit(), Qubit(), Qubit()) {
            // Prepare target qubit in a zero- or one-state, based on input value
            if (inputValue) {
                X(target);
            }
            PrepareQubit(inputBasis, target);

            within {
                H(auxiliary);
                H(resource);
            }
            apply {
                ApplyRzArcTan2(limit, auxiliary, resource, target);
                // let rotation angle = 2.0 * ArcTan(2.0);
                let rotationAngle = 
                2.2142974355881810060341309203570740801400952908028652933530784148;
                Rz(rotationAngle, target); // Rotate back to initial state
            }
            return [M(auxiliary), M(resource), Measure([measurementBasis], [target])];
        }
    }

    /// # Summary
    /// Apply Rz(arctan(2)) on qubits using repeat until success algorithm.
    operation ApplyRzArcTan2 (
        limit : Int, 
        auxiliary : Qubit, 
        resource : Qubit, 
        target : Qubit
    )
    : Unit {

        // Run Part 1 of the program.
        let result1 = ApplyAndMeasurePart1(auxiliary, resource);
        // We'll only run Part 2 if Part 1 returns Zero.
        // Otherwise, we'll skip and rerun Part 1 again.
        if (result1 == Zero) {
            let result2 = ApplyAndMeasurePart2(resource, target);
            if (result2 == One and limit > 1) {
                Z(resource); // Reset resource
                Adjoint Z(target); // Correct effective Z rotation on target
                // Run next iteration
                ApplyRzArcTan2(limit - 1, auxiliary, resource, target);
            }
        } elif (limit > 1) {
            Z(auxiliary);
            Reset(resource);
            H(resource);
            ApplyRzArcTan2(limit - 1, auxiliary, resource, target);
        }
    }

    /// # Summary
    /// Apply part 1 of RUS circuit and measure
    /// auxiliary qubit in Pauli X basis
    operation ApplyAndMeasurePart1(auxiliary : Qubit, resource : Qubit) : Result {
        within {
            T(auxiliary);
        } apply {
            CNOT(resource, auxiliary);
        }

        return Measure([PauliX], [auxiliary]);
    }

    /// # Summary
    /// Apply part 2 of RUS circuit and measure
    /// resource qubit in Pauli X basis
    operation ApplyAndMeasurePart2(resource : Qubit, target : Qubit)  : Result {
        T(target);
        Z(target);
        CNOT(target, resource);
        T(resource);

        return Measure([PauliX], [resource]);
    }
}
\end{lstlisting}
\begin{lstlisting}[label={lst:qsharp-rus-loop},language=qsharp,caption={\qsharp~program implementing the circuit in \autoref{fig:RUS_cir} using a bounded loop. All ommitted operations match \autoref{lst:qsharp-rus-recursion}},escapeinside={(<}{>)},basicstyle = \ttfamily\tiny]
    /// # Summary
    /// Apply Rz(arctan(2)) on qubits using repeat until success algorithm.
    /// Updated to use for-loop sytle.
    operation ApplyRzArcTan2 (
        limit : Int, 
        auxiliary : Qubit, 
        resource : Qubit, 
        target : Qubit
    )
    : Unit {

        mutable (result1, result2) = (false, false);
        // Run Part 1 of the program.
        for _ in 1..limit {
            if not result1 or not result2 {
                set result1 = ApplyAndMeasurePart1(auxiliary, resource) == Zero;
                // We'll only run Part 2 if Part 1 returns Zero.
                // Otherwise, we'll skip and rerun Part 1 again.
                if result1 {
                    set result2 = ApplyAndMeasurePart2(resource, target) == Zero;
                    if not result2 {
                        Z(resource); // Reset resource
                        Adjoint Z(target); // Correct effective Z rotation on target
                    }
                } else {
                    Z(auxiliary);
                    Reset(resource);
                    H(resource);
                }
            }
        }
    }
\end{lstlisting}

%%%%%%%%%%%%%%%%%%%%%%%%%%%%%%%%%%%%%%%%%%%%%%%%%%%%%%%%%%%%%%%%%%%%%%%%%%%%%%
\section{QIR Examples}
\label{apx:qir-examples}
%%%%%%%%%%%%%%%%%%%%%%%%%%%%%%%%%%%%%%%%%%%%%%%%%%%%%%%%%%%%%%%%%%%%%%%%%%%%%%

\begin{lstlisting}[caption={Snippet of QIR generated from \autoref{alg:rus} with limit 1 in Pauli-Y basis for the for loop implementation.},gobble=4,mathescape=false,basicstyle = \ttfamily\tiny,escapeinside={;<}{;>},breaklines=true,label={lst:rus-qir}]
    ; ModuleID = 'trueYY1.bc'
    source_filename = "qat-link"
    
    %Qubit = type opaque
    %Result = type opaque
    
    define void @Microsoft__Quantum__Samples__AllocateQubitsAndApplyRzArcTan2__main() #0 {
    entry:
      call void @__quantum__qis__x__body(%Qubit* null)
      call void @__quantum__qis__h__body(%Qubit* null)
      call void @__quantum__qis__s__body(%Qubit* null)
      call void @__quantum__qis__h__body(%Qubit* nonnull inttoptr (i64 1 to %Qubit*))
      call void @__quantum__qis__h__body(%Qubit* nonnull inttoptr (i64 2 to %Qubit*))
      call void @__quantum__qis__t__body(%Qubit* nonnull inttoptr (i64 1 to %Qubit*))
      call void @__quantum__qis__cnot__body(%Qubit* nonnull inttoptr (i64 2 to %Qubit*), %Qubit* nonnull inttoptr (i64 1 to %Qubit*))
      call void @__quantum__qis__t__adj(%Qubit* nonnull inttoptr (i64 1 to %Qubit*))
      call void @__quantum__qis__h__body(%Qubit* nonnull inttoptr (i64 1 to %Qubit*))
      call void @__quantum__qis__mz__body(%Qubit* nonnull inttoptr (i64 1 to %Qubit*), %Result* null)
      call void @__quantum__qis__reset__body(%Qubit* nonnull inttoptr (i64 1 to %Qubit*))
      %0 = call i1 @__quantum__qis__read_result__body(%Result* null)
      br i1 %0, label %then0__1, label %continue__6
    
    then0__1:                                         ; preds = %entry
      call void @__quantum__qis__x__body(%Qubit* nonnull inttoptr (i64 1 to %Qubit*))
      br label %continue__6
    
    continue__6:                                      ; preds = %then0__1, %entry
      call void @__quantum__qis__h__body(%Qubit* nonnull inttoptr (i64 1 to %Qubit*))
      %1 = call i1 @__quantum__qis__read_result__body(%Result* null)
      br i1 %1, label %continue__8, label %then0__2
    
    then0__2:                                         ; preds = %continue__6
      call void @__quantum__qis__t__body(%Qubit* null)
      call void @__quantum__qis__z__body(%Qubit* null)
      call void @__quantum__qis__cnot__body(%Qubit* null, %Qubit* nonnull inttoptr (i64 2 to %Qubit*))
      call void @__quantum__qis__t__body(%Qubit* nonnull inttoptr (i64 2 to %Qubit*))
      call void @__quantum__qis__h__body(%Qubit* nonnull inttoptr (i64 2 to %Qubit*))
      call void @__quantum__qis__mz__body(%Qubit* nonnull inttoptr (i64 2 to %Qubit*), %Result* nonnull inttoptr (i64 1 to %Result*))
      call void @__quantum__qis__reset__body(%Qubit* nonnull inttoptr (i64 2 to %Qubit*))
      %2 = call i1 @__quantum__qis__read_result__body(%Result* nonnull inttoptr (i64 1 to %Result*))
      br i1 %2, label %then0__3, label %continue__12
    
    then0__3:                                         ; preds = %then0__2
      call void @__quantum__qis__x__body(%Qubit* nonnull inttoptr (i64 2 to %Qubit*))
      br label %continue__12
    
    continue__12:                                     ; preds = %then0__3, %then0__2
      call void @__quantum__qis__h__body(%Qubit* nonnull inttoptr (i64 2 to %Qubit*))
      %3 = call i1 @__quantum__qis__read_result__body(%Result* nonnull inttoptr (i64 1 to %Result*))
      br label %continue__8
    
    continue__8:                                      ; preds = %continue__12, %continue__6
      call void @__quantum__qis__rz__body(double 0x4001B6E192EBBE44, %Qubit* null)
      call void @__quantum__qis__h__body(%Qubit* nonnull inttoptr (i64 2 to %Qubit*))
      call void @__quantum__qis__h__body(%Qubit* nonnull inttoptr (i64 1 to %Qubit*))
      call void @__quantum__qis__mz__body(%Qubit* nonnull inttoptr (i64 1 to %Qubit*), %Result* nonnull inttoptr (i64 2 to %Result*))
      call void @__quantum__qis__reset__body(%Qubit* nonnull inttoptr (i64 1 to %Qubit*))
      %4 = call i1 @__quantum__qis__read_result__body(%Result* nonnull inttoptr (i64 2 to %Result*))
      br i1 %4, label %then0__4, label %continue__15
    
    then0__4:                                         ; preds = %continue__8
      call void @__quantum__qis__x__body(%Qubit* nonnull inttoptr (i64 1 to %Qubit*))
      br label %continue__15
    
    continue__15:                                     ; preds = %then0__4, %continue__8
      call void @__quantum__qis__mz__body(%Qubit* nonnull inttoptr (i64 2 to %Qubit*), %Result* nonnull inttoptr (i64 3 to %Result*))
      call void @__quantum__qis__reset__body(%Qubit* nonnull inttoptr (i64 2 to %Qubit*))
      %5 = call i1 @__quantum__qis__read_result__body(%Result* nonnull inttoptr (i64 3 to %Result*))
      br i1 %5, label %then0__5, label %continue__16
    
    then0__5:                                         ; preds = %continue__15
      call void @__quantum__qis__x__body(%Qubit* nonnull inttoptr (i64 2 to %Qubit*))
      br label %continue__16
    
    continue__16:                                     ; preds = %then0__5, %continue__15
      call void @__quantum__qis__h__body(%Qubit* null)
      call void @__quantum__qis__s__body(%Qubit* null)
      call void @__quantum__qis__h__body(%Qubit* null)
      call void @__quantum__qis__mz__body(%Qubit* null, %Result* nonnull inttoptr (i64 4 to %Result*))
      call void @__quantum__qis__reset__body(%Qubit* null)
      %6 = call i1 @__quantum__qis__read_result__body(%Result* nonnull inttoptr (i64 4 to %Result*))
      br i1 %6, label %then0__6, label %continue__20
    
    then0__6:                                         ; preds = %continue__16
      call void @__quantum__qis__x__body(%Qubit* null)
      br label %continue__20
    
    continue__20:                                     ; preds = %then0__6, %continue__16
      call void @__quantum__qis__h__body(%Qubit* null)
      call void @__quantum__qis__s__adj(%Qubit* null)
      call void @__quantum__qis__h__body(%Qubit* null)
      call void @__quantum__rt__array_start_record_output()
      call void @__quantum__rt__result_record_output(%Result* nonnull inttoptr (i64 2 to %Result*))
      call void @__quantum__rt__result_record_output(%Result* nonnull inttoptr (i64 3 to %Result*))
      call void @__quantum__rt__result_record_output(%Result* nonnull inttoptr (i64 4 to %Result*))
      call void @__quantum__rt__array_end_record_output()
      ret void
    }
    
    declare void @__quantum__qis__x__body(%Qubit*)
    
    declare void @__quantum__qis__h__body(%Qubit*)
    
    declare void @__quantum__qis__s__body(%Qubit*)
    
    declare void @__quantum__qis__t__body(%Qubit*)
    
    declare void @__quantum__qis__cnot__body(%Qubit*, %Qubit*)
    
    declare void @__quantum__qis__t__adj(%Qubit*)
    
    declare void @__quantum__qis__reset__body(%Qubit*)
    
    declare void @__quantum__qis__z__body(%Qubit*)
    
    declare void @__quantum__qis__rz__body(double, %Qubit*)
    
    declare void @__quantum__qis__s__adj(%Qubit*)
    
    declare void @__quantum__rt__array_start_record_output()
    
    declare void @__quantum__rt__result_record_output(%Result*)
    
    declare void @__quantum__rt__array_end_record_output()
    
    declare void @__quantum__qis__mz__body(%Qubit*, %Result*)
    
    declare i1 @__quantum__qis__read_result__body(%Result*)
    
    attributes #0 = { "EntryPoint" "maxQubitIndex"="2" "maxResultIndex"="4" "requiredQubits"="3" "requiredResults"="5" }
\end{lstlisting}

%%%%%%%%%%%%%%%%%%%%%%%%%%%%%%%%%%%%%%%%%%%%%%%%%%%%%%%%%%%%%%%%%%%%%%%%%%%%%%
\section{Repeat-until-success representations}
\label{apx:rus-examples}
%%%%%%%%%%%%%%%%%%%%%%%%%%%%%%%%%%%%%%%%%%%%%%%%%%%%%%%%%%%%%%%%%%%%%%%%%%%%%%

\begin{lstlisting}[caption={Handwritten OpenQASM2.0++ version of the RUS circuit at limit = 3},label={ext_qasm},basicstyle = \ttfamily\tiny]
OPENQASM 2.0;
include "qelib1.inc";
qreg q[3];
creg c0[1];
creg c1[1];
creg c2[1];

c1 = 1;  //OpenQASM extension
//Prepare
if(c1==1) reset q[2]; //OpenQASM extension
//if(c1==1) h q[2]; for X basis 
//if(c1==1) h q[2]; for Y basis
//if(c1==1) s q[2]; for Y basis 
if(c1==1) t q[2];
if(c1==1) z q[2];
if(c1==1) reset q[0]; //OpenQASM extension
if(c1==1) reset q[1]; //OpenQASM extension
if(c1==1) h q[0];
if(c1==1) h q[1];

//First part of circuit
if(c1==1) tdg q[0];
if(c1==1) cx q[1],q[0];
if(c1==1) t q[0];
if(c1==1) h q[0];
if(c1==1) measure q[0] -> c0[0]; //OpenQASM extension

//Second part of circuit
if(c0==0) cx q[2],q[1];
if(c0==0) t q[1];
if(c0==0) h q[1];
if(c0==0) measure q[1] -> c1[0]; //OpenQASM extension
if(c1==0) c0 = 1; //OpenQASM extension

//Prepare
if(c1==1) reset q[2]; //OpenQASM extension
//if(c1==1) h q[2]; for X basis 
//if(c1==1) h q[2]; for Y basis
//if(c1==1) s q[2]; for Y basis 
if(c1==1) t q[2];
if(c1==1) z q[2];
if(c1==1) reset q[0]; //OpenQASM extension
if(c1==1) reset q[1]; //OpenQASM extension
if(c1==1) h q[0];
if(c1==1) h q[1];

//First part of circuit
if(c1==1) tdg q[0];
if(c1==1) cx q[1],q[0];
if(c1==1) t q[0];
if(c1==1) h q[0];
if(c1==1) measure q[0] -> c0[0]; //OpenQASM extension

//Second part of circuit
if(c0==0) cx q[2],q[1];
if(c0==0) t q[1];
if(c0==0) h q[1];
if(c0==0) measure q[1] -> c1[0]; //OpenQASM extension
if(c1==0) c0 = 1; //OpenQASM extension

//Prepare
if(c1==1) reset q[2]; //OpenQASM extension
//if(c1==1) h q[2]; for X basis 
//if(c1==1) h q[2]; for Y basis
//if(c1==1) s q[2]; for Y basis 
if(c1==1) t q[2];
if(c1==1) z q[2];
if(c1==1) reset q[0]; //OpenQASM extension
if(c1==1) reset q[1]; //OpenQASM extension
if(c1==1) h q[0];
if(c1==1) h q[1];

//First part of circuit
if(c1==1) tdg q[0];
if(c1==1) cx q[1],q[0];
if(c1==1) t q[0];
if(c1==1) h q[0];
if(c1==1) measure q[0] -> c0[0]; //OpenQASM extension

//Second part of circuit
if(c0==0) cx q[2],q[1];
if(c0==0) t q[1];
if(c0==0) h q[1];
if(c0==0) measure q[1] -> c1[0]; //OpenQASM extension
if(c1==0) c0 = 1; //OpenQASM extension

//final measure out 
rz(2.2142974) q[2];
//h q[2]; for X basis
// sdg q[2]; for Y basis
// h q[2]; for Y basis 
measure q[2] -> c2[0];
\end{lstlisting}

\begin{lstlisting}[caption={ $Q^ \#$ to OpenQASM2.0++ - recursion version of the RUS circuit at limit = 3, Z basis},label={Q_qasm},basicstyle = \ttfamily\tiny]
OPENQASM 2.0;
include "qelib1.inc";

qreg q[3];
creg c0[6];
creg c1[17];

h q[0];
h q[1];
t q[0];
cx q[1], q[0];
tdg q[0];
h q[0];
measure q[0] -> c1[0];
reset q[0];
c0[0] = c1[0]; //OpenQASM extension
if (c0==1) x q[0];
c0[0] = 0; //OpenQASM extension
h q[0];
c0[0] = c1[0]; //OpenQASM extension
if (c0==0) t q[2];
if (c0==0) z q[2];
if (c0==0) cx q[2], q[1];
if (c0==0) t q[1];
if (c0==0) h q[1];
if (c0==0) measure q[1] -> c1[1]; //OpenQASM extension
if (c0==0) reset q[1]; //OpenQASM extension
c0[1] = c1[1]; //OpenQASM extension
if (c0==2) x q[1];
c0[1] = 0; //OpenQASM extension
if (c0==0) h q[1];
c0[1] = c1[1]; //OpenQASM extension
if (c0==2) z q[1];
if (c0==2) z q[2];
if (c0==2) t q[0];
if (c0==2) cx q[1], q[0];
if (c0==2) tdg q[0];
if (c0==2) h q[0];
if (c0==2) measure q[0] -> c1[2]; //OpenQASM extension
if (c0==2) reset q[0]; //OpenQASM extension
c0[2] = c1[2]; //OpenQASM extension
if (c0==6) x q[0];
c0[2] = 0; //OpenQASM extension
if (c0==2) h q[0];
c0[2] = c1[2]; //OpenQASM extension
if (c0==2) t q[2];
if (c0==2) z q[2];
if (c0==2) cx q[2], q[1];
if (c0==2) t q[1];
if (c0==2) h q[1];
if (c0==2) measure q[1] -> c1[3]; //OpenQASM extension
if (c0==2) reset q[1]; //OpenQASM extension
c0[3] = c1[3]; //OpenQASM extension
if (c0==10) x q[1];
c0[3] = 0; //OpenQASM extension
if (c0==2) h q[1];
c0[3] = c1[3]; //OpenQASM extension
if (c0==10) z q[1];
if (c0==10) z q[2];
if (c0==10) t q[0];
if (c0==10) cx q[1], q[0];
if (c0==10) tdg q[0];
if (c0==10) h q[0];
if (c0==10) measure q[0] -> c1[4]; //OpenQASM extension
if (c0==10) reset q[0]; //OpenQASM extension
c0[4] = c1[4]; //OpenQASM extension
if (c0==26) x q[0];
c0[4] = 0; //OpenQASM extension
if (c0==10) h q[0];
c0[4] = c1[4]; //OpenQASM extension
if (c0==10) t q[2];
if (c0==10) z q[2];
if (c0==10) cx q[2], q[1];
if (c0==10) t q[1];
if (c0==10) h q[1];
if (c0==10) measure q[1] -> c1[5]; //OpenQASM extension
if (c0==10) reset q[1]; //OpenQASM extension
c0[5] = c1[5]; //OpenQASM extension
if (c0==42) x q[1];
c0[5] = 0; //OpenQASM extension
if (c0==10) h q[1];
c0[4] = 0; //OpenQASM extension
c0[3] = 0; //OpenQASM extension
c0[2] = 0; //OpenQASM extension
c0[2] = c1[2]; //OpenQASM extension
if (c0==6) z q[0];
if (c0==6) reset q[1]; //OpenQASM extension
if (c0==6) h q[1];
if (c0==6) t q[0];
if (c0==6) cx q[1], q[0];
if (c0==6) tdg q[0];
if (c0==6) h q[0];
if (c0==6) measure q[0] -> c1[6]; //OpenQASM extension
if (c0==6) reset q[0]; //OpenQASM extension
c0[3] = c1[6]; //OpenQASM extension
if (c0==14) x q[0];
c0[3] = 0; //OpenQASM extension
if (c0==6) h q[0];
c0[3] = c1[6]; //OpenQASM extension
if (c0==6) t q[2];
if (c0==6) z q[2];
if (c0==6) cx q[2], q[1];
if (c0==6) t q[1];
if (c0==6) h q[1];
if (c0==6) measure q[1] -> c1[7]; //OpenQASM extension
if (c0==6) reset q[1]; //OpenQASM extension
c0[4] = c1[7]; //OpenQASM extension
if (c0==22) x q[1];
c0[4] = 0; //OpenQASM extension
if (c0==6) h q[1];
c0[3] = 0; //OpenQASM extension
c0[2] = 0; //OpenQASM extension
c0[1] = 0; //OpenQASM extension
c0[0] = 0; //OpenQASM extension
c0[0] = c1[0]; //OpenQASM extension
if (c0==1) z q[0];
if (c0==1) reset q[1]; //OpenQASM extension
if (c0==1) h q[1];
if (c0==1) t q[0];
if (c0==1) cx q[1], q[0];
if (c0==1) tdg q[0];
if (c0==1) h q[0];
if (c0==1) measure q[0] -> c1[8]; //OpenQASM extension
if (c0==1) reset q[0]; //OpenQASM extension
c0[1] = c1[8]; //OpenQASM extension
if (c0==3) x q[0];
c0[1] = 0; //OpenQASM extension
if (c0==1) h q[0];
c0[1] = c1[8]; //OpenQASM extension
if (c0==1) t q[2];
if (c0==1) z q[2];
if (c0==1) cx q[2], q[1];
if (c0==1) t q[1];
if (c0==1) h q[1];
if (c0==1) measure q[1] -> c1[9]; //OpenQASM extension
if (c0==1) reset q[1]; //OpenQASM extension
c0[2] = c1[9]; //OpenQASM extension
if (c0==5) x q[1];
c0[2] = 0; //OpenQASM extension
if (c0==1) h q[1];
c0[2] = c1[9]; //OpenQASM extension
if (c0==5) z q[1];
if (c0==5) z q[2];
if (c0==5) t q[0];
if (c0==5) cx q[1], q[0];
if (c0==5) tdg q[0];
if (c0==5) h q[0];
if (c0==5) measure q[0] -> c1[10]; //OpenQASM extension
if (c0==5) reset q[0]; //OpenQASM extension
c0[3] = c1[10]; //OpenQASM extension
if (c0==13) x q[0];
c0[3] = 0; //OpenQASM extension
if (c0==5) h q[0];
c0[3] = c1[10]; //OpenQASM extension
if (c0==5) t q[2];
if (c0==5) z q[2];
if (c0==5) cx q[2], q[1];
if (c0==5) t q[1];
if (c0==5) h q[1];
if (c0==5) measure q[1] -> c1[11]; //OpenQASM extension
if (c0==5) reset q[1]; //OpenQASM extension
c0[4] = c1[11]; //OpenQASM extension
if (c0==21) x q[1];
c0[4] = 0; //OpenQASM extension
if (c0==5) h q[1];
c0[3] = 0; //OpenQASM extension
c0[2] = 0; //OpenQASM extension
c0[1] = 0; //OpenQASM extension
c0[1] = c1[8]; //OpenQASM extension
if (c0==3) z q[0];
if (c0==3) reset q[1]; //OpenQASM extension
if (c0==3) h q[1];
if (c0==3) t q[0];
if (c0==3) cx q[1], q[0];
if (c0==3) tdg q[0];
if (c0==3) h q[0];
if (c0==3) measure q[0] -> c1[12]; //OpenQASM extension
if (c0==3) reset q[0]; //OpenQASM extension
c0[2] = c1[12]; //OpenQASM extension
if (c0==7) x q[0];
c0[2] = 0; //OpenQASM extension
if (c0==3) h q[0];
c0[2] = c1[12]; //OpenQASM extension
if (c0==3) t q[2];
if (c0==3) z q[2];
if (c0==3) cx q[2], q[1];
if (c0==3) t q[1];
if (c0==3) h q[1];
if (c0==3) measure q[1] -> c1[13]; //OpenQASM extension
if (c0==3) reset q[1]; //OpenQASM extension
c0[3] = c1[13]; //OpenQASM extension
if (c0==11) x q[1];
c0[3] = 0; //OpenQASM extension
if (c0==3) h q[1];
c0[2] = 0; //OpenQASM extension
c0[1] = 0; //OpenQASM extension
c0[0] = 0; //OpenQASM extension
rz(2.214297435588181) q[2];
h q[1];
h q[0];
measure q[0] -> c1[14];
measure q[1] -> c1[15];
measure q[2] -> c1[16];
\end{lstlisting}

\begin{lstlisting}[caption={ $Q^ \#$ to QIR - recursion version of the RUS circuit at limit = 3, Z basis},label={Q_qirrecursion},basicstyle = \ttfamily\tiny]
  ; ModuleID = 'limit3.bc'
source_filename = "qat-link"

%Qubit = type opaque
%Result = type opaque
%String = type opaque

@0 = internal constant [29 x i8] c"limiting recursion depth to \00"
@1 = internal constant [13 x i8] c"inputValue: \00"
@2 = internal constant [6 x i8] c"false\00"
@3 = internal constant [15 x i8] c", inputBasis: \00"
@4 = internal constant [7 x i8] c"PauliZ\00"
@5 = internal constant [21 x i8] c", measurementBasis: \00"

define void @Microsoft__Quantum__Qir__Development__RunExample() #0 {
entry:
  call void @__quantum__qis__h__body(%Qubit* null)
  call void @__quantum__qis__h__body(%Qubit* nonnull inttoptr (i64 1 to %Qubit*))
  call void @__quantum__qis__t__body(%Qubit* null)
  call void @__quantum__qis__cnot__body(%Qubit* nonnull inttoptr (i64 1 to %Qubit*), %Qubit* null)
  call void @__quantum__qis__t__adj(%Qubit* null)
  call void @__quantum__qis__h__body(%Qubit* null)
  call void @__quantum__qis__mz__body(%Qubit* null, %Result* null)
  call void @__quantum__qis__reset__body(%Qubit* null)
  %0 = call i1 @__quantum__qis__read_result__body(%Result* null)
  br i1 %0, label %then0__1, label %continue__6

then0__1:                                         ; preds = %entry
  call void @__quantum__qis__x__body(%Qubit* null)
  br label %continue__6

continue__6:                                      ; preds = %then0__1, %entry
  call void @__quantum__qis__h__body(%Qubit* null)
  %1 = call i1 @__quantum__qis__read_result__body(%Result* null)
  br i1 %1, label %test1__1, label %then0__2

then0__2:                                         ; preds = %continue__6
  call void @__quantum__qis__t__body(%Qubit* nonnull inttoptr (i64 2 to %Qubit*))
  call void @__quantum__qis__z__body(%Qubit* nonnull inttoptr (i64 2 to %Qubit*))
  call void @__quantum__qis__cnot__body(%Qubit* nonnull inttoptr (i64 2 to %Qubit*), %Qubit* nonnull inttoptr (i64 1 to %Qubit*))
  call void @__quantum__qis__t__body(%Qubit* nonnull inttoptr (i64 1 to %Qubit*))
  call void @__quantum__qis__h__body(%Qubit* nonnull inttoptr (i64 1 to %Qubit*))
  call void @__quantum__qis__mz__body(%Qubit* nonnull inttoptr (i64 1 to %Qubit*), %Result* nonnull inttoptr (i64 1 to %Result*))
  call void @__quantum__qis__reset__body(%Qubit* nonnull inttoptr (i64 1 to %Qubit*))
  %2 = call i1 @__quantum__qis__read_result__body(%Result* nonnull inttoptr (i64 1 to %Result*))
  br i1 %2, label %then0__3, label %continue__12

then0__3:                                         ; preds = %then0__2
  call void @__quantum__qis__x__body(%Qubit* nonnull inttoptr (i64 1 to %Qubit*))
  br label %continue__12

continue__12:                                     ; preds = %then0__3, %then0__2
  call void @__quantum__qis__h__body(%Qubit* nonnull inttoptr (i64 1 to %Qubit*))
  %3 = call i1 @__quantum__qis__read_result__body(%Result* nonnull inttoptr (i64 1 to %Result*))
  br i1 %3, label %then0__4, label %continue__8

then0__4:                                         ; preds = %continue__12
  call void @__quantum__qis__z__body(%Qubit* nonnull inttoptr (i64 1 to %Qubit*))
  call void @__quantum__qis__z__body(%Qubit* nonnull inttoptr (i64 2 to %Qubit*))
  call void @__quantum__qis__t__body(%Qubit* null)
  call void @__quantum__qis__cnot__body(%Qubit* nonnull inttoptr (i64 1 to %Qubit*), %Qubit* null)
  call void @__quantum__qis__t__adj(%Qubit* null)
  call void @__quantum__qis__h__body(%Qubit* null)
  call void @__quantum__qis__mz__body(%Qubit* null, %Result* nonnull inttoptr (i64 2 to %Result*))
  call void @__quantum__qis__reset__body(%Qubit* null)
  %4 = call i1 @__quantum__qis__read_result__body(%Result* nonnull inttoptr (i64 2 to %Result*))
  br i1 %4, label %then0__5, label %continue__18

then0__5:                                         ; preds = %then0__4
  call void @__quantum__qis__x__body(%Qubit* null)
  br label %continue__18

continue__18:                                     ; preds = %then0__5, %then0__4
  call void @__quantum__qis__h__body(%Qubit* null)
  %5 = call i1 @__quantum__qis__read_result__body(%Result* nonnull inttoptr (i64 2 to %Result*))
  br i1 %5, label %test1__2, label %then0__6

then0__6:                                         ; preds = %continue__18
  call void @__quantum__qis__t__body(%Qubit* nonnull inttoptr (i64 2 to %Qubit*))
  call void @__quantum__qis__z__body(%Qubit* nonnull inttoptr (i64 2 to %Qubit*))
  call void @__quantum__qis__cnot__body(%Qubit* nonnull inttoptr (i64 2 to %Qubit*), %Qubit* nonnull inttoptr (i64 1 to %Qubit*))
  call void @__quantum__qis__t__body(%Qubit* nonnull inttoptr (i64 1 to %Qubit*))
  call void @__quantum__qis__h__body(%Qubit* nonnull inttoptr (i64 1 to %Qubit*))
  call void @__quantum__qis__mz__body(%Qubit* nonnull inttoptr (i64 1 to %Qubit*), %Result* nonnull inttoptr (i64 3 to %Result*))
  call void @__quantum__qis__reset__body(%Qubit* nonnull inttoptr (i64 1 to %Qubit*))
  %6 = call i1 @__quantum__qis__read_result__body(%Result* nonnull inttoptr (i64 3 to %Result*))
  br i1 %6, label %then0__7, label %continue__24

then0__7:                                         ; preds = %then0__6
  call void @__quantum__qis__x__body(%Qubit* nonnull inttoptr (i64 1 to %Qubit*))
  br label %continue__24

continue__24:                                     ; preds = %then0__7, %then0__6
  call void @__quantum__qis__h__body(%Qubit* nonnull inttoptr (i64 1 to %Qubit*))
  %7 = call i1 @__quantum__qis__read_result__body(%Result* nonnull inttoptr (i64 3 to %Result*))
  br i1 %7, label %then0__8, label %continue__8

then0__8:                                         ; preds = %continue__24
  call void @__quantum__qis__z__body(%Qubit* nonnull inttoptr (i64 1 to %Qubit*))
  call void @__quantum__qis__z__body(%Qubit* nonnull inttoptr (i64 2 to %Qubit*))
  call void @__quantum__qis__t__body(%Qubit* null)
  call void @__quantum__qis__cnot__body(%Qubit* nonnull inttoptr (i64 1 to %Qubit*), %Qubit* null)
  call void @__quantum__qis__t__adj(%Qubit* null)
  call void @__quantum__qis__h__body(%Qubit* null)
  call void @__quantum__qis__mz__body(%Qubit* null, %Result* nonnull inttoptr (i64 4 to %Result*))
  call void @__quantum__qis__reset__body(%Qubit* null)
  %8 = call i1 @__quantum__qis__read_result__body(%Result* nonnull inttoptr (i64 4 to %Result*))
  br i1 %8, label %then0__9, label %continue__30

then0__9:                                         ; preds = %then0__8
  call void @__quantum__qis__x__body(%Qubit* null)
  br label %continue__30

continue__30:                                     ; preds = %then0__9, %then0__8
  call void @__quantum__qis__h__body(%Qubit* null)
  %9 = call i1 @__quantum__qis__read_result__body(%Result* nonnull inttoptr (i64 4 to %Result*))
  br i1 %9, label %continue__8, label %then0__10

then0__10:                                        ; preds = %continue__30
  call void @__quantum__qis__t__body(%Qubit* nonnull inttoptr (i64 2 to %Qubit*))
  call void @__quantum__qis__z__body(%Qubit* nonnull inttoptr (i64 2 to %Qubit*))
  call void @__quantum__qis__cnot__body(%Qubit* nonnull inttoptr (i64 2 to %Qubit*), %Qubit* nonnull inttoptr (i64 1 to %Qubit*))
  call void @__quantum__qis__t__body(%Qubit* nonnull inttoptr (i64 1 to %Qubit*))
  call void @__quantum__qis__h__body(%Qubit* nonnull inttoptr (i64 1 to %Qubit*))
  call void @__quantum__qis__mz__body(%Qubit* nonnull inttoptr (i64 1 to %Qubit*), %Result* nonnull inttoptr (i64 5 to %Result*))
  call void @__quantum__qis__reset__body(%Qubit* nonnull inttoptr (i64 1 to %Qubit*))
  %10 = call i1 @__quantum__qis__read_result__body(%Result* nonnull inttoptr (i64 5 to %Result*))
  br i1 %10, label %then0__11, label %continue__36

then0__11:                                        ; preds = %then0__10
  call void @__quantum__qis__x__body(%Qubit* nonnull inttoptr (i64 1 to %Qubit*))
  br label %continue__36

continue__36:                                     ; preds = %then0__11, %then0__10
  call void @__quantum__qis__h__body(%Qubit* nonnull inttoptr (i64 1 to %Qubit*))
  %11 = call i1 @__quantum__qis__read_result__body(%Result* nonnull inttoptr (i64 5 to %Result*))
  br label %continue__8

test1__2:                                         ; preds = %continue__18
  call void @__quantum__qis__z__body(%Qubit* null)
  call void @__quantum__qis__reset__body(%Qubit* nonnull inttoptr (i64 1 to %Qubit*))
  call void @__quantum__qis__h__body(%Qubit* nonnull inttoptr (i64 1 to %Qubit*))
  call void @__quantum__qis__t__body(%Qubit* null)
  call void @__quantum__qis__cnot__body(%Qubit* nonnull inttoptr (i64 1 to %Qubit*), %Qubit* null)
  call void @__quantum__qis__t__adj(%Qubit* null)
  call void @__quantum__qis__h__body(%Qubit* null)
  call void @__quantum__qis__mz__body(%Qubit* null, %Result* nonnull inttoptr (i64 6 to %Result*))
  call void @__quantum__qis__reset__body(%Qubit* null)
  %12 = call i1 @__quantum__qis__read_result__body(%Result* nonnull inttoptr (i64 6 to %Result*))
  br i1 %12, label %then0__12, label %continue__42

then0__12:                                        ; preds = %test1__2
  call void @__quantum__qis__x__body(%Qubit* null)
  br label %continue__42

continue__42:                                     ; preds = %then0__12, %test1__2
  call void @__quantum__qis__h__body(%Qubit* null)
  %13 = call i1 @__quantum__qis__read_result__body(%Result* nonnull inttoptr (i64 6 to %Result*))
  br i1 %13, label %continue__8, label %then0__13

then0__13:                                        ; preds = %continue__42
  call void @__quantum__qis__t__body(%Qubit* nonnull inttoptr (i64 2 to %Qubit*))
  call void @__quantum__qis__z__body(%Qubit* nonnull inttoptr (i64 2 to %Qubit*))
  call void @__quantum__qis__cnot__body(%Qubit* nonnull inttoptr (i64 2 to %Qubit*), %Qubit* nonnull inttoptr (i64 1 to %Qubit*))
  call void @__quantum__qis__t__body(%Qubit* nonnull inttoptr (i64 1 to %Qubit*))
  call void @__quantum__qis__h__body(%Qubit* nonnull inttoptr (i64 1 to %Qubit*))
  call void @__quantum__qis__mz__body(%Qubit* nonnull inttoptr (i64 1 to %Qubit*), %Result* nonnull inttoptr (i64 7 to %Result*))
  call void @__quantum__qis__reset__body(%Qubit* nonnull inttoptr (i64 1 to %Qubit*))
  %14 = call i1 @__quantum__qis__read_result__body(%Result* nonnull inttoptr (i64 7 to %Result*))
  br i1 %14, label %then0__14, label %continue__48

then0__14:                                        ; preds = %then0__13
  call void @__quantum__qis__x__body(%Qubit* nonnull inttoptr (i64 1 to %Qubit*))
  br label %continue__48

continue__48:                                     ; preds = %then0__14, %then0__13
  call void @__quantum__qis__h__body(%Qubit* nonnull inttoptr (i64 1 to %Qubit*))
  %15 = call i1 @__quantum__qis__read_result__body(%Result* nonnull inttoptr (i64 7 to %Result*))
  br label %continue__8

test1__1:                                         ; preds = %continue__6
  call void @__quantum__qis__z__body(%Qubit* null)
  call void @__quantum__qis__reset__body(%Qubit* nonnull inttoptr (i64 1 to %Qubit*))
  call void @__quantum__qis__h__body(%Qubit* nonnull inttoptr (i64 1 to %Qubit*))
  call void @__quantum__qis__t__body(%Qubit* null)
  call void @__quantum__qis__cnot__body(%Qubit* nonnull inttoptr (i64 1 to %Qubit*), %Qubit* null)
  call void @__quantum__qis__t__adj(%Qubit* null)
  call void @__quantum__qis__h__body(%Qubit* null)
  call void @__quantum__qis__mz__body(%Qubit* null, %Result* nonnull inttoptr (i64 8 to %Result*))
  call void @__quantum__qis__reset__body(%Qubit* null)
  %16 = call i1 @__quantum__qis__read_result__body(%Result* nonnull inttoptr (i64 8 to %Result*))
  br i1 %16, label %then0__15, label %continue__54

then0__15:                                        ; preds = %test1__1
  call void @__quantum__qis__x__body(%Qubit* null)
  br label %continue__54

continue__54:                                     ; preds = %then0__15, %test1__1
  call void @__quantum__qis__h__body(%Qubit* null)
  %17 = call i1 @__quantum__qis__read_result__body(%Result* nonnull inttoptr (i64 8 to %Result*))
  br i1 %17, label %test1__5, label %then0__16

then0__16:                                        ; preds = %continue__54
  call void @__quantum__qis__t__body(%Qubit* nonnull inttoptr (i64 2 to %Qubit*))
  call void @__quantum__qis__z__body(%Qubit* nonnull inttoptr (i64 2 to %Qubit*))
  call void @__quantum__qis__cnot__body(%Qubit* nonnull inttoptr (i64 2 to %Qubit*), %Qubit* nonnull inttoptr (i64 1 to %Qubit*))
  call void @__quantum__qis__t__body(%Qubit* nonnull inttoptr (i64 1 to %Qubit*))
  call void @__quantum__qis__h__body(%Qubit* nonnull inttoptr (i64 1 to %Qubit*))
  call void @__quantum__qis__mz__body(%Qubit* nonnull inttoptr (i64 1 to %Qubit*), %Result* nonnull inttoptr (i64 9 to %Result*))
  call void @__quantum__qis__reset__body(%Qubit* nonnull inttoptr (i64 1 to %Qubit*))
  %18 = call i1 @__quantum__qis__read_result__body(%Result* nonnull inttoptr (i64 9 to %Result*))
  br i1 %18, label %then0__17, label %continue__60

then0__17:                                        ; preds = %then0__16
  call void @__quantum__qis__x__body(%Qubit* nonnull inttoptr (i64 1 to %Qubit*))
  br label %continue__60

continue__60:                                     ; preds = %then0__17, %then0__16
  call void @__quantum__qis__h__body(%Qubit* nonnull inttoptr (i64 1 to %Qubit*))
  %19 = call i1 @__quantum__qis__read_result__body(%Result* nonnull inttoptr (i64 9 to %Result*))
  br i1 %19, label %then0__18, label %continue__8

then0__18:                                        ; preds = %continue__60
  call void @__quantum__qis__z__body(%Qubit* nonnull inttoptr (i64 1 to %Qubit*))
  call void @__quantum__qis__z__body(%Qubit* nonnull inttoptr (i64 2 to %Qubit*))
  call void @__quantum__qis__t__body(%Qubit* null)
  call void @__quantum__qis__cnot__body(%Qubit* nonnull inttoptr (i64 1 to %Qubit*), %Qubit* null)
  call void @__quantum__qis__t__adj(%Qubit* null)
  call void @__quantum__qis__h__body(%Qubit* null)
  call void @__quantum__qis__mz__body(%Qubit* null, %Result* nonnull inttoptr (i64 10 to %Result*))
  call void @__quantum__qis__reset__body(%Qubit* null)
  %20 = call i1 @__quantum__qis__read_result__body(%Result* nonnull inttoptr (i64 10 to %Result*))
  br i1 %20, label %then0__19, label %continue__66

then0__19:                                        ; preds = %then0__18
  call void @__quantum__qis__x__body(%Qubit* null)
  br label %continue__66

continue__66:                                     ; preds = %then0__19, %then0__18
  call void @__quantum__qis__h__body(%Qubit* null)
  %21 = call i1 @__quantum__qis__read_result__body(%Result* nonnull inttoptr (i64 10 to %Result*))
  br i1 %21, label %continue__8, label %then0__20

then0__20:                                        ; preds = %continue__66
  call void @__quantum__qis__t__body(%Qubit* nonnull inttoptr (i64 2 to %Qubit*))
  call void @__quantum__qis__z__body(%Qubit* nonnull inttoptr (i64 2 to %Qubit*))
  call void @__quantum__qis__cnot__body(%Qubit* nonnull inttoptr (i64 2 to %Qubit*), %Qubit* nonnull inttoptr (i64 1 to %Qubit*))
  call void @__quantum__qis__t__body(%Qubit* nonnull inttoptr (i64 1 to %Qubit*))
  call void @__quantum__qis__h__body(%Qubit* nonnull inttoptr (i64 1 to %Qubit*))
  call void @__quantum__qis__mz__body(%Qubit* nonnull inttoptr (i64 1 to %Qubit*), %Result* nonnull inttoptr (i64 11 to %Result*))
  call void @__quantum__qis__reset__body(%Qubit* nonnull inttoptr (i64 1 to %Qubit*))
  %22 = call i1 @__quantum__qis__read_result__body(%Result* nonnull inttoptr (i64 11 to %Result*))
  br i1 %22, label %then0__21, label %continue__72

then0__21:                                        ; preds = %then0__20
  call void @__quantum__qis__x__body(%Qubit* nonnull inttoptr (i64 1 to %Qubit*))
  br label %continue__72

continue__72:                                     ; preds = %then0__21, %then0__20
  call void @__quantum__qis__h__body(%Qubit* nonnull inttoptr (i64 1 to %Qubit*))
  %23 = call i1 @__quantum__qis__read_result__body(%Result* nonnull inttoptr (i64 11 to %Result*))
  br label %continue__8

test1__5:                                         ; preds = %continue__54
  call void @__quantum__qis__z__body(%Qubit* null)
  call void @__quantum__qis__reset__body(%Qubit* nonnull inttoptr (i64 1 to %Qubit*))
  call void @__quantum__qis__h__body(%Qubit* nonnull inttoptr (i64 1 to %Qubit*))
  call void @__quantum__qis__t__body(%Qubit* null)
  call void @__quantum__qis__cnot__body(%Qubit* nonnull inttoptr (i64 1 to %Qubit*), %Qubit* null)
  call void @__quantum__qis__t__adj(%Qubit* null)
  call void @__quantum__qis__h__body(%Qubit* null)
  call void @__quantum__qis__mz__body(%Qubit* null, %Result* nonnull inttoptr (i64 12 to %Result*))
  call void @__quantum__qis__reset__body(%Qubit* null)
  %24 = call i1 @__quantum__qis__read_result__body(%Result* nonnull inttoptr (i64 12 to %Result*))
  br i1 %24, label %then0__22, label %continue__78

then0__22:                                        ; preds = %test1__5
  call void @__quantum__qis__x__body(%Qubit* null)
  br label %continue__78

continue__78:                                     ; preds = %then0__22, %test1__5
  call void @__quantum__qis__h__body(%Qubit* null)
  %25 = call i1 @__quantum__qis__read_result__body(%Result* nonnull inttoptr (i64 12 to %Result*))
  br i1 %25, label %continue__8, label %then0__23

then0__23:                                        ; preds = %continue__78
  call void @__quantum__qis__t__body(%Qubit* nonnull inttoptr (i64 2 to %Qubit*))
  call void @__quantum__qis__z__body(%Qubit* nonnull inttoptr (i64 2 to %Qubit*))
  call void @__quantum__qis__cnot__body(%Qubit* nonnull inttoptr (i64 2 to %Qubit*), %Qubit* nonnull inttoptr (i64 1 to %Qubit*))
  call void @__quantum__qis__t__body(%Qubit* nonnull inttoptr (i64 1 to %Qubit*))
  call void @__quantum__qis__h__body(%Qubit* nonnull inttoptr (i64 1 to %Qubit*))
  call void @__quantum__qis__mz__body(%Qubit* nonnull inttoptr (i64 1 to %Qubit*), %Result* nonnull inttoptr (i64 13 to %Result*))
  call void @__quantum__qis__reset__body(%Qubit* nonnull inttoptr (i64 1 to %Qubit*))
  %26 = call i1 @__quantum__qis__read_result__body(%Result* nonnull inttoptr (i64 13 to %Result*))
  br i1 %26, label %then0__24, label %continue__84

then0__24:                                        ; preds = %then0__23
  call void @__quantum__qis__x__body(%Qubit* nonnull inttoptr (i64 1 to %Qubit*))
  br label %continue__84

continue__84:                                     ; preds = %then0__24, %then0__23
  call void @__quantum__qis__h__body(%Qubit* nonnull inttoptr (i64 1 to %Qubit*))
  %27 = call i1 @__quantum__qis__read_result__body(%Result* nonnull inttoptr (i64 13 to %Result*))
  br label %continue__8

continue__8:                                      ; preds = %continue__84, %continue__78, %continue__72, %continue__66, %continue__60, %continue__48, %continue__42, %continue__36, %continue__30, %continue__24, %continue__12
  call void @__quantum__qis__rz__body(double 0x4001B6E192EBBE44, %Qubit* nonnull inttoptr (i64 2 to %Qubit*))
  call void @__quantum__qis__h__body(%Qubit* nonnull inttoptr (i64 1 to %Qubit*))
  call void @__quantum__qis__h__body(%Qubit* null)
  call void @__quantum__qis__mz__body(%Qubit* null, %Result* nonnull inttoptr (i64 14 to %Result*))
  call void @__quantum__qis__reset__body(%Qubit* null)
  %28 = call i1 @__quantum__qis__read_result__body(%Result* nonnull inttoptr (i64 14 to %Result*))
  br i1 %28, label %then0__25, label %continue__87

then0__25:                                        ; preds = %continue__8
  call void @__quantum__qis__x__body(%Qubit* null)
  br label %continue__87

continue__87:                                     ; preds = %then0__25, %continue__8
  call void @__quantum__qis__mz__body(%Qubit* nonnull inttoptr (i64 1 to %Qubit*), %Result* nonnull inttoptr (i64 15 to %Result*))
  call void @__quantum__qis__reset__body(%Qubit* nonnull inttoptr (i64 1 to %Qubit*))
  %29 = call i1 @__quantum__qis__read_result__body(%Result* nonnull inttoptr (i64 15 to %Result*))
  br i1 %29, label %then0__26, label %continue__88

then0__26:                                        ; preds = %continue__87
  call void @__quantum__qis__x__body(%Qubit* nonnull inttoptr (i64 1 to %Qubit*))
  br label %continue__88

continue__88:                                     ; preds = %then0__26, %continue__87
  call void @__quantum__qis__mz__body(%Qubit* nonnull inttoptr (i64 2 to %Qubit*), %Result* nonnull inttoptr (i64 16 to %Result*))
  call void @__quantum__qis__reset__body(%Qubit* nonnull inttoptr (i64 2 to %Qubit*))
  %30 = call i1 @__quantum__qis__read_result__body(%Result* nonnull inttoptr (i64 16 to %Result*))
  br i1 %30, label %then0__27, label %continue__92

then0__27:                                        ; preds = %continue__88
  call void @__quantum__qis__x__body(%Qubit* nonnull inttoptr (i64 2 to %Qubit*))
  br label %continue__92

continue__92:                                     ; preds = %then0__27, %continue__88
  call void @__quantum__rt__array_start_record_output()
  call void @__quantum__rt__result_record_output(%Result* nonnull inttoptr (i64 14 to %Result*))
  call void @__quantum__rt__result_record_output(%Result* nonnull inttoptr (i64 15 to %Result*))
  call void @__quantum__rt__result_record_output(%Result* nonnull inttoptr (i64 16 to %Result*))
  call void @__quantum__rt__array_end_record_output()
  ret void
}

declare %String* @__quantum__rt__string_create(i8*)

declare %String* @__quantum__rt__int_to_string(i64)

declare %String* @__quantum__rt__string_concatenate(%String*, %String*)

declare void @__quantum__rt__string_update_reference_count(%String*, i32)

declare void @__quantum__rt__message(%String*)

declare %Qubit* @__quantum__rt__qubit_allocate()

declare void @__quantum__qis__h__body(%Qubit*)

declare void @__quantum__qis__t__body(%Qubit*)

declare void @__quantum__qis__cnot__body(%Qubit*, %Qubit*)

declare void @__quantum__qis__t__adj(%Qubit*)

declare %Result* @__quantum__rt__result_get_zero()

declare void @__quantum__rt__result_update_reference_count(%Result*, i32)

declare %Result* @__quantum__qis__m__body(%Qubit*)

declare void @__quantum__qis__reset__body(%Qubit*)

declare %Result* @__quantum__rt__result_get_one()

declare i1 @__quantum__rt__result_equal(%Result*, %Result*)

declare void @__quantum__qis__x__body(%Qubit*)

declare void @__quantum__qis__z__body(%Qubit*)

declare void @__quantum__qis__rz__body(double, %Qubit*)

declare void @__quantum__rt__qubit_release(%Qubit*)

declare void @__quantum__rt__array_start_record_output()

declare void @__quantum__rt__result_record_output(%Result*)

declare void @__quantum__rt__array_end_record_output()

declare void @__quantum__qis__mz__body(%Qubit*, %Result*)

declare i1 @__quantum__qis__read_result__body(%Result*)

attributes #0 = { "EntryPoint" "maxQubitIndex"="2" "maxResultIndex"="16" "requiredQubits"="3" "requiredResults"="17" }
\end{lstlisting}

\begin{lstlisting}[caption={ $Q^ \#$ to QIR - loop version of the RUS circuit at limit = 3, Z basis},label={Q_qirloop}, basicstyle = \ttfamily\tiny]
; ModuleID = 'RUSLoopZZ-3.bc'
source_filename = "qat-link"

%Qubit = type opaque
%Result = type opaque

@0 = internal constant [6 x i8] c"0_t0r\00"
@1 = internal constant [6 x i8] c"1_t1r\00"
@2 = internal constant [6 x i8] c"2_t2r\00"

define void @Microsoft__Quantum__Samples__RepeatUntilSuccess__RepeatUntilSuccess() #0 {
entry:
  call void @__quantum__qis__h__body(%Qubit* null)
  call void @__quantum__qis__h__body(%Qubit* nonnull inttoptr (i64 1 to %Qubit*))
  call void @__quantum__qis__t__body(%Qubit* null)
  call void @__quantum__qis__cnot__body(%Qubit* nonnull inttoptr (i64 1 to %Qubit*), %Qubit* null)
  call void @__quantum__qis__t__adj(%Qubit* null)
  call void @__quantum__qis__h__body(%Qubit* null)
  call void @__quantum__qis__mz__body(%Qubit* null, %Result* null)
  call void @__quantum__qis__reset__body(%Qubit* null)
  %0 = call i1 @__quantum__qis__read_result__body(%Result* null)
  br i1 %0, label %then0__3, label %continue__7

then0__3:                                         ; preds = %entry
  call void @__quantum__qis__x__body(%Qubit* null)
  br label %continue__7

continue__7:                                      ; preds = %then0__3, %entry
  call void @__quantum__qis__h__body(%Qubit* null)
  %1 = call i1 @__quantum__qis__read_result__body(%Result* null)
  br i1 %1, label %else__1, label %then0__4

then0__4:                                         ; preds = %continue__7
  call void @__quantum__qis__t__body(%Qubit* nonnull inttoptr (i64 2 to %Qubit*))
  call void @__quantum__qis__z__body(%Qubit* nonnull inttoptr (i64 2 to %Qubit*))
  call void @__quantum__qis__cnot__body(%Qubit* nonnull inttoptr (i64 2 to %Qubit*), %Qubit* nonnull inttoptr (i64 1 to %Qubit*))
  call void @__quantum__qis__t__body(%Qubit* nonnull inttoptr (i64 1 to %Qubit*))
  call void @__quantum__qis__h__body(%Qubit* nonnull inttoptr (i64 1 to %Qubit*))
  call void @__quantum__qis__mz__body(%Qubit* nonnull inttoptr (i64 1 to %Qubit*), %Result* nonnull inttoptr (i64 1 to %Result*))
  call void @__quantum__qis__reset__body(%Qubit* nonnull inttoptr (i64 1 to %Qubit*))
  %2 = call i1 @__quantum__qis__read_result__body(%Result* nonnull inttoptr (i64 1 to %Result*))
  br i1 %2, label %then0__6, label %continue__13

then0__6:                                         ; preds = %then0__4
  call void @__quantum__qis__x__body(%Qubit* nonnull inttoptr (i64 1 to %Qubit*))
  br label %continue__13

continue__13:                                     ; preds = %then0__6, %then0__4
  call void @__quantum__qis__h__body(%Qubit* nonnull inttoptr (i64 1 to %Qubit*))
  %3 = call i1 @__quantum__qis__read_result__body(%Result* nonnull inttoptr (i64 1 to %Result*))
  %4 = xor i1 %3, true
  br i1 %3, label %then0__7, label %continue__9

then0__7:                                         ; preds = %continue__13
  call void @__quantum__qis__z__body(%Qubit* nonnull inttoptr (i64 1 to %Qubit*))
  call void @__quantum__qis__z__body(%Qubit* nonnull inttoptr (i64 2 to %Qubit*))
  br label %continue__9

else__1:                                          ; preds = %continue__7
  call void @__quantum__qis__z__body(%Qubit* null)
  call void @__quantum__qis__reset__body(%Qubit* nonnull inttoptr (i64 1 to %Qubit*))
  call void @__quantum__qis__h__body(%Qubit* nonnull inttoptr (i64 1 to %Qubit*))
  br label %continue__9

continue__9:                                      ; preds = %else__1, %then0__7, %continue__13
  %result2.0 = phi i1 [ false, %else__1 ], [ %4, %then0__7 ], [ %4, %continue__13 ]
  %5 = xor i1 %result2.0, true
  %6 = select i1 %1, i1 true, i1 %5
  br i1 %6, label %then0__8, label %continue__16

then0__8:                                         ; preds = %continue__9
  call void @__quantum__qis__t__body(%Qubit* null)
  call void @__quantum__qis__cnot__body(%Qubit* nonnull inttoptr (i64 1 to %Qubit*), %Qubit* null)
  call void @__quantum__qis__t__adj(%Qubit* null)
  call void @__quantum__qis__h__body(%Qubit* null)
  call void @__quantum__qis__mz__body(%Qubit* null, %Result* nonnull inttoptr (i64 2 to %Result*))
  call void @__quantum__qis__reset__body(%Qubit* null)
  %7 = call i1 @__quantum__qis__read_result__body(%Result* nonnull inttoptr (i64 2 to %Result*))
  br i1 %7, label %then0__10, label %continue__20

then0__10:                                        ; preds = %then0__8
  call void @__quantum__qis__x__body(%Qubit* null)
  br label %continue__20

continue__20:                                     ; preds = %then0__10, %then0__8
  call void @__quantum__qis__h__body(%Qubit* null)
  %8 = call i1 @__quantum__qis__read_result__body(%Result* nonnull inttoptr (i64 2 to %Result*))
  br i1 %8, label %else__2, label %then0__11

then0__11:                                        ; preds = %continue__20
  call void @__quantum__qis__t__body(%Qubit* nonnull inttoptr (i64 2 to %Qubit*))
  call void @__quantum__qis__z__body(%Qubit* nonnull inttoptr (i64 2 to %Qubit*))
  call void @__quantum__qis__cnot__body(%Qubit* nonnull inttoptr (i64 2 to %Qubit*), %Qubit* nonnull inttoptr (i64 1 to %Qubit*))
  call void @__quantum__qis__t__body(%Qubit* nonnull inttoptr (i64 1 to %Qubit*))
  call void @__quantum__qis__h__body(%Qubit* nonnull inttoptr (i64 1 to %Qubit*))
  call void @__quantum__qis__mz__body(%Qubit* nonnull inttoptr (i64 1 to %Qubit*), %Result* nonnull inttoptr (i64 3 to %Result*))
  call void @__quantum__qis__reset__body(%Qubit* nonnull inttoptr (i64 1 to %Qubit*))
  %9 = call i1 @__quantum__qis__read_result__body(%Result* nonnull inttoptr (i64 3 to %Result*))
  br i1 %9, label %then0__13, label %continue__26

then0__13:                                        ; preds = %then0__11
  call void @__quantum__qis__x__body(%Qubit* nonnull inttoptr (i64 1 to %Qubit*))
  br label %continue__26

continue__26:                                     ; preds = %then0__13, %then0__11
  call void @__quantum__qis__h__body(%Qubit* nonnull inttoptr (i64 1 to %Qubit*))
  %10 = call i1 @__quantum__qis__read_result__body(%Result* nonnull inttoptr (i64 3 to %Result*))
  %11 = xor i1 %10, true
  br i1 %10, label %then0__14, label %continue__16

then0__14:                                        ; preds = %continue__26
  call void @__quantum__qis__z__body(%Qubit* nonnull inttoptr (i64 1 to %Qubit*))
  call void @__quantum__qis__z__body(%Qubit* nonnull inttoptr (i64 2 to %Qubit*))
  br label %continue__16

else__2:                                          ; preds = %continue__20
  call void @__quantum__qis__z__body(%Qubit* null)
  call void @__quantum__qis__reset__body(%Qubit* nonnull inttoptr (i64 1 to %Qubit*))
  call void @__quantum__qis__h__body(%Qubit* nonnull inttoptr (i64 1 to %Qubit*))
  br label %continue__16

continue__16:                                     ; preds = %else__2, %then0__14, %continue__26, %continue__9
  %result1.0.in = phi i1 [ %1, %continue__9 ], [ %8, %continue__26 ], [ %8, %then0__14 ], [ %8, %else__2 ]
  %result2.2 = phi i1 [ %result2.0, %continue__9 ], [ %11, %continue__26 ], [ %11, %then0__14 ], [ %result2.0, %else__2 ]
  %result1.0 = xor i1 %result1.0.in, true
  %12 = select i1 %result1.0, i1 %result2.2, i1 false
  br i1 %12, label %continue__29, label %then0__15

then0__15:                                        ; preds = %continue__16
  call void @__quantum__qis__t__body(%Qubit* null)
  call void @__quantum__qis__cnot__body(%Qubit* nonnull inttoptr (i64 1 to %Qubit*), %Qubit* null)
  call void @__quantum__qis__t__adj(%Qubit* null)
  call void @__quantum__qis__h__body(%Qubit* null)
  call void @__quantum__qis__mz__body(%Qubit* null, %Result* nonnull inttoptr (i64 4 to %Result*))
  call void @__quantum__qis__reset__body(%Qubit* null)
  %13 = call i1 @__quantum__qis__read_result__body(%Result* nonnull inttoptr (i64 4 to %Result*))
  br i1 %13, label %then0__17, label %continue__33

then0__17:                                        ; preds = %then0__15
  call void @__quantum__qis__x__body(%Qubit* null)
  br label %continue__33

continue__33:                                     ; preds = %then0__17, %then0__15
  call void @__quantum__qis__h__body(%Qubit* null)
  %14 = call i1 @__quantum__qis__read_result__body(%Result* nonnull inttoptr (i64 4 to %Result*))
  br i1 %14, label %else__3, label %then0__18

then0__18:                                        ; preds = %continue__33
  call void @__quantum__qis__t__body(%Qubit* nonnull inttoptr (i64 2 to %Qubit*))
  call void @__quantum__qis__z__body(%Qubit* nonnull inttoptr (i64 2 to %Qubit*))
  call void @__quantum__qis__cnot__body(%Qubit* nonnull inttoptr (i64 2 to %Qubit*), %Qubit* nonnull inttoptr (i64 1 to %Qubit*))
  call void @__quantum__qis__t__body(%Qubit* nonnull inttoptr (i64 1 to %Qubit*))
  call void @__quantum__qis__h__body(%Qubit* nonnull inttoptr (i64 1 to %Qubit*))
  call void @__quantum__qis__mz__body(%Qubit* nonnull inttoptr (i64 1 to %Qubit*), %Result* nonnull inttoptr (i64 5 to %Result*))
  call void @__quantum__qis__reset__body(%Qubit* nonnull inttoptr (i64 1 to %Qubit*))
  %15 = call i1 @__quantum__qis__read_result__body(%Result* nonnull inttoptr (i64 5 to %Result*))
  br i1 %15, label %then0__20, label %continue__39

then0__20:                                        ; preds = %then0__18
  call void @__quantum__qis__x__body(%Qubit* nonnull inttoptr (i64 1 to %Qubit*))
  br label %continue__39

continue__39:                                     ; preds = %then0__20, %then0__18
  call void @__quantum__qis__h__body(%Qubit* nonnull inttoptr (i64 1 to %Qubit*))
  %16 = call i1 @__quantum__qis__read_result__body(%Result* nonnull inttoptr (i64 5 to %Result*))
  br i1 %16, label %then0__21, label %continue__29

then0__21:                                        ; preds = %continue__39
  call void @__quantum__qis__z__body(%Qubit* nonnull inttoptr (i64 1 to %Qubit*))
  call void @__quantum__qis__z__body(%Qubit* nonnull inttoptr (i64 2 to %Qubit*))
  br label %continue__29

else__3:                                          ; preds = %continue__33
  call void @__quantum__qis__z__body(%Qubit* null)
  call void @__quantum__qis__reset__body(%Qubit* nonnull inttoptr (i64 1 to %Qubit*))
  call void @__quantum__qis__h__body(%Qubit* nonnull inttoptr (i64 1 to %Qubit*))
  br label %continue__29

continue__29:                                     ; preds = %else__3, %then0__21, %continue__39, %continue__16
  call void @__quantum__qis__rz__body(double 0x4001B6E192EBBE42, %Qubit* nonnull inttoptr (i64 2 to %Qubit*))
  call void @__quantum__qis__h__body(%Qubit* nonnull inttoptr (i64 1 to %Qubit*))
  call void @__quantum__qis__h__body(%Qubit* null)
  call void @__quantum__qis__mz__body(%Qubit* null, %Result* nonnull inttoptr (i64 6 to %Result*))
  call void @__quantum__qis__reset__body(%Qubit* null)
  call void @__quantum__qis__mz__body(%Qubit* nonnull inttoptr (i64 1 to %Qubit*), %Result* nonnull inttoptr (i64 7 to %Result*))
  call void @__quantum__qis__reset__body(%Qubit* nonnull inttoptr (i64 1 to %Qubit*))
  call void @__quantum__qis__mz__body(%Qubit* nonnull inttoptr (i64 2 to %Qubit*), %Result* nonnull inttoptr (i64 8 to %Result*))
  call void @__quantum__qis__reset__body(%Qubit* nonnull inttoptr (i64 2 to %Qubit*))
  %17 = call i1 @__quantum__qis__read_result__body(%Result* nonnull inttoptr (i64 8 to %Result*))
  br i1 %17, label %then0__24, label %continue__45

then0__24:                                        ; preds = %continue__29
  call void @__quantum__qis__x__body(%Qubit* nonnull inttoptr (i64 2 to %Qubit*))
  br label %continue__45

continue__45:                                     ; preds = %then0__24, %continue__29
  call void @__quantum__rt__tuple_start_record_output()
  call void @__quantum__rt__result_record_output(%Result* nonnull inttoptr (i64 6 to %Result*), i8* getelementptr inbounds ([6 x i8], [6 x i8]* @0, i64 0, i64 0))
  call void @__quantum__rt__result_record_output(%Result* nonnull inttoptr (i64 7 to %Result*), i8* getelementptr inbounds ([6 x i8], [6 x i8]* @1, i64 0, i64 0))
  call void @__quantum__rt__result_record_output(%Result* nonnull inttoptr (i64 8 to %Result*), i8* getelementptr inbounds ([6 x i8], [6 x i8]* @2, i64 0, i64 0))
  call void @__quantum__rt__tuple_end_record_output()
  ret void
}

declare %Qubit* @__quantum__rt__qubit_allocate()

declare void @__quantum__qis__h__body(%Qubit*)

declare void @__quantum__qis__t__body(%Qubit*)

declare void @__quantum__qis__cnot__body(%Qubit*, %Qubit*)

declare void @__quantum__qis__t__adj(%Qubit*)

declare %Result* @__quantum__rt__result_get_zero()

declare void @__quantum__rt__result_update_reference_count(%Result*, i32)

declare %Result* @__quantum__qis__m__body(%Qubit*)

declare void @__quantum__qis__reset__body(%Qubit*)

declare %Result* @__quantum__rt__result_get_one()

declare i1 @__quantum__rt__result_equal(%Result*, %Result*)

declare void @__quantum__qis__x__body(%Qubit*)

declare void @__quantum__qis__z__body(%Qubit*)

declare void @__quantum__qis__rz__body(double, %Qubit*)

declare void @__quantum__rt__qubit_release(%Qubit*)

declare void @__quantum__rt__tuple_start_record_output()

declare void @__quantum__rt__result_record_output(%Result*, i8*)

declare void @__quantum__rt__tuple_end_record_output()

declare void @__quantum__qis__mz__body(%Qubit*, %Result*)

declare i1 @__quantum__qis__read_result__body(%Result*)

attributes #0 = { "EntryPoint" "maxQubitIndex"="2" "maxResultIndex"="8" "requiredQubits"="3" "requiredResults"="9" }
\end{lstlisting}

\newpage
\section{Control flow graphs}

\begin{figure}[h!]
%trim=left bottom right top, clip
\begin{center}
\includegraphics[trim=0 0 0 0, clip, scale=1.10]{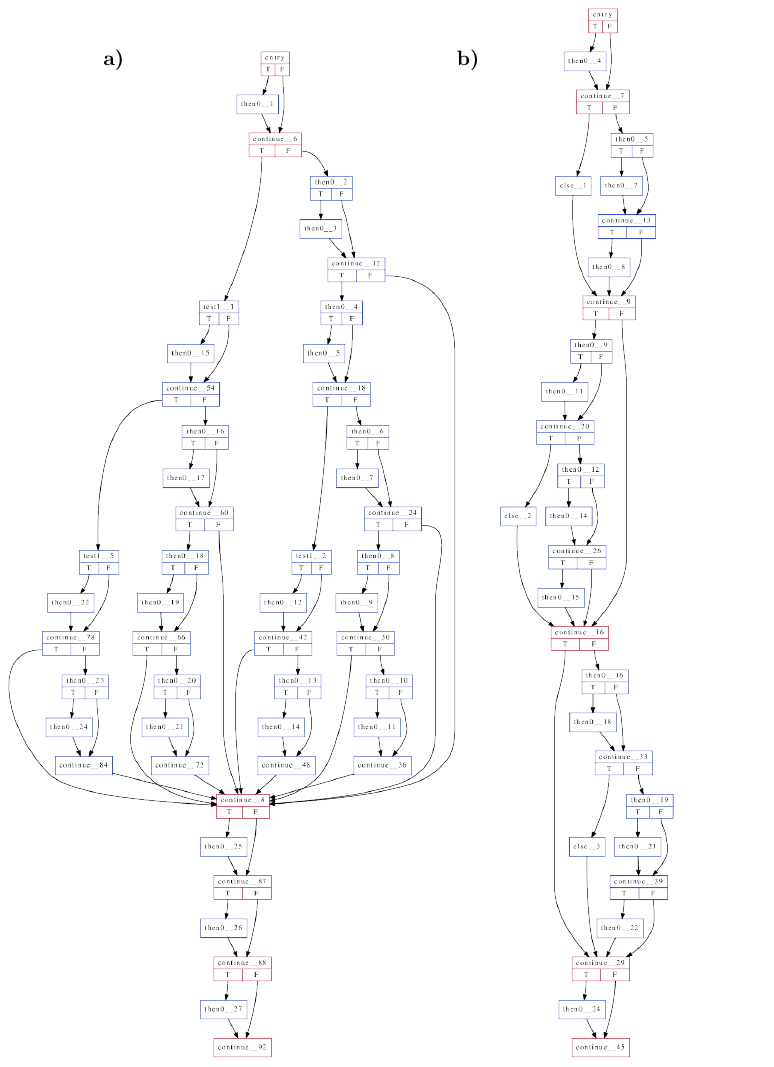}
  \caption{Control flow graphs for the RUS circuit for both the recursion implementation \textbf{a)}, and the for loop implementation \textbf{b)}. The recursive looping structure admits a more complex tree structure, ulimately leading to an increase in the number of transport steps needed to implement the circuit.}
  \label{apx:RUS_CFG}
\end{center}
\end{figure}

\section{Additional Simulations}

\begin{figure}[h!]
\includegraphics[trim=0 0 0 0, clip,width=\textwidth]{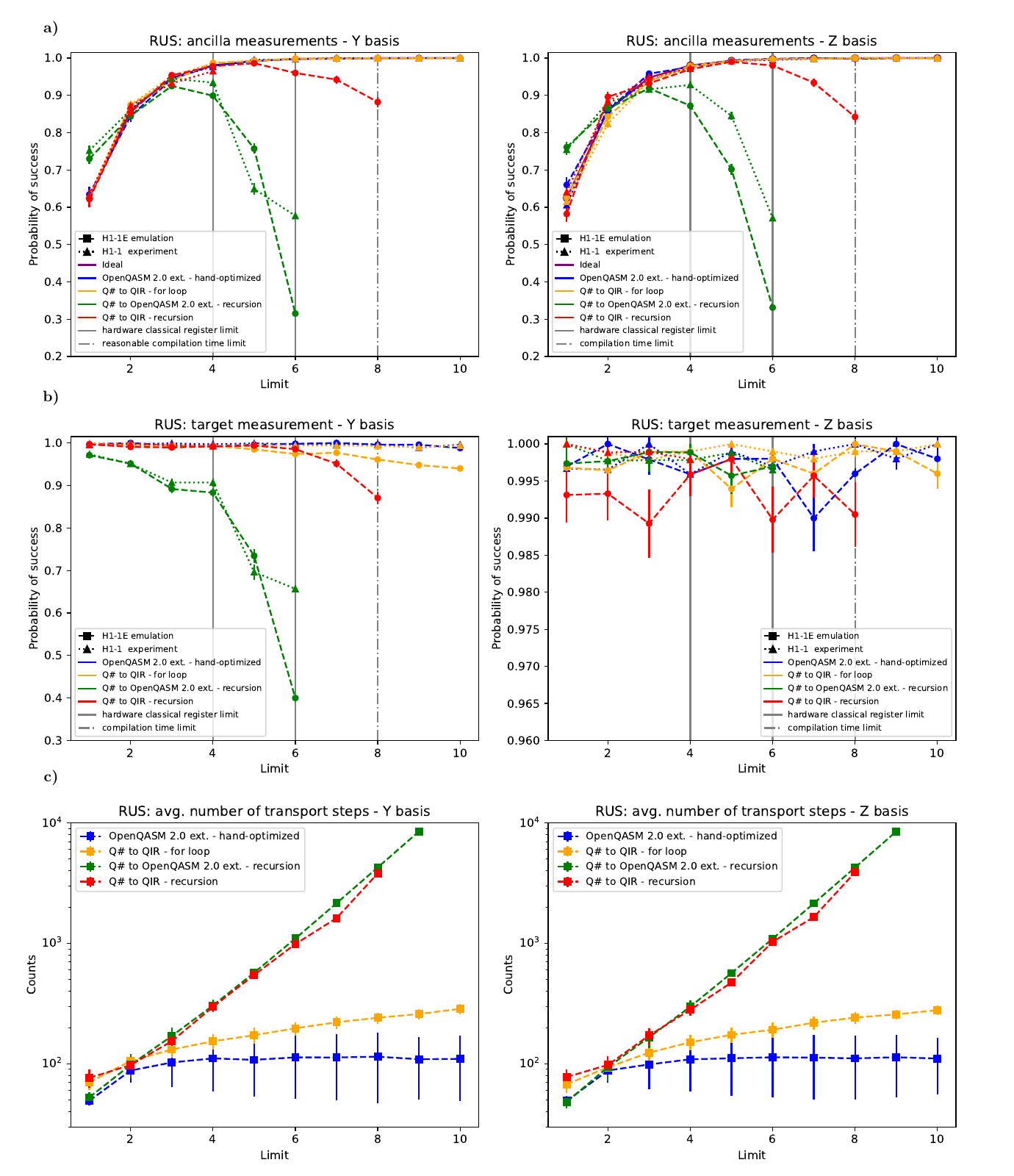}
\caption{Additional RUS circuits for Pauli $Y$ and $Z$ bases. The way in which transport noise is modeled as has a greater impact in the $Y$ basis, thus we see a bigger impact on performance for the $Y$ basis over the $Z$ basis. Further explaination of why the number of transport steps scales with respect to the different implementations is discussed in the main text. More details of the simulation model can be found in \cite{ryan2021realization, ryan2022implementing}}
\label{RUS_YZ}
\end{figure}

\end{document}